# Physical Layer Security of Autonomous Driving: Secure Vehicle-to-Vehicle Communication in A Security Cluster

Na-Young Ahn and Dong Hoon Lee

*Abstract*—We suggest secure Vehicle-to-Vehicle communications in a secure cluster. Here, the security cluster refers to a group of vehicles having a certain level or more of secrecy capacity. Usually, there are many difficulties in defining secrecy capacity, but we define vehicular secrecy capacity for the vehicle defined only by SNR values. Defined vehicular secrecy capacity is practical and efficient in achieving physical layer security in V2V. Typically, secrecy capacity may be changed by antenna related parameters, path related parameters, and noise related parameters. In addition to these conventional parameters, we address unique vehicle-related parameters, such as vehicle speed, safety distance, speed limit, response time, etc. in connection with autonomous driving. We confirm the relationship between vehicle-related secrecy parameters and secrecy capacity through modeling in highway and urban traffic situations. These vehicular secrecy parameters enable real-time control of vehicle secrecy capacity of V2V communications. We can use vehicular secrecy capacity to achieve secure vehicle communications from attackers such as quantum computers. Our research enables economic, effective and efficient physical layer security in autonomous driving.

*Index Terms*—**Vehicle-to-Vehicle, Secrecy Capacity, Physical Layer Security, Autonomous Driving, Vehicle Speed, Safety Distance, Vehicular Secrecy Capacity, vehicle-related secrecy parameters, Quantum Computer**

## I. Introduction

In the near future, it will be mandatory for all vehicles to be equipped with a vehicle-to-vehicle (V2V) communication functionality. To this end, the National Highway Traffic Safety Administration (NHTSA) even recently published a proposition that would require all new manufactured vehicles to have these capabilities. Although the law has not yet passed, manufacturers may begin to phase this technology into their fleets, when, by law, all models will be required to feature V2V [1]. Aside from vehicles, V2V functionality is essential for general 5G communications [2–4]. This functionality aims to achieve the safe operation of autonomous vehicles by sharing vehicle driving-related information, such as basic security messages (BSMs). To guarantee V2V functionality, security must be the foundation of design and privacy must be a priority. Previous studies have already made significant progress on security beyond the vehicular network layer [5–6]. Yet, despite this, existing vehicle security methods demonstrate insufficient computing power and large power consumption with respect to processing received or transmitted data from a large number of vehicles. To overcome these difficulties, studies on physical layer security [7–9] have attempted to develop secure data communication methods based on the physical properties of the radio channel in the wireless communications field. The basic idea of physical layer security is that noise and fading can be used to efficiently hide data from attackers without sacrificing significant data rates. Unfortunately, relevant research in the vehicular communication field is still scarce concerning this area. In addition, the emergence of quantum computers will disrupt traditional cryptographic communication schemes, which will increasingly require the need for physical layer security in wireless communications.

A recent study suggests the possibility of secure and efficient V2V communications using a secrecy capacity, which is informally defined as the data rate of confidential data [10]. They investigated secrecy capacity factors (e.g., vehicle speed, response time, and transmission power) that are limited to vehicle communication, confirming that security parameters can be controlled by using these parameters. Following their approach [10], this study explores the means to ensure secure vehicular communications by exploiting the physical layer security. Existing studies have attempted to calculate secrecy capacity by modeling the system but these efforts have failed to provide meaningful information concerning actual vehicle communication. The most difficult problem is how to define the model of eavesdroppers. As mentioned in [10], we now know what types of eavesdroppers exist in real environments. This study makes a novel contribution by reducing the gap between these real environments and theory. We define secrecy capacity of the vehicle with only Signal-to-Noise Ratio (SNR) values provided in existing wireless communications to perform vehicle communication using the vehicle's defined secrecy capacity. In particular, we propose secure vehicle communication within a security cluster defined by the vehicle secrecy capacity. We expect secure vehicle communications over 5G using a simple security cluster algorithm while communicating over existing Multi-In-Multi-Out (MIMO) antennas.

In section II, the vehicle-specific security parameters required to achieve physical layer security in V2V communications are presented. In section III, modeling of highway conditions and urban traffic conditions was performed between vehicular secrecy parameters and the secrecy capacity presented in V2V communications. In section IV, the general secrecy capacity is presented and vehicular secrecy capacity for the vehicle for vehicle communications is introduced. In section V, various embodiments of V2V communication using vehicular secrecy capacity are introduced. Basically, proposed V2V communication used a security cluster. In section VI, we introduced compression sensing techniques, data encryption using MEC to improve secrecy capacity. In section VIII, simulation results for highway conditions and urban traffic conditions are presented.

II. SAFETY DISTANCE, SPEED LIMIT, AND SECRECY CAPACITY

Unlike conventional wireless communications, we have primarily identified factors that would inherently impact secrecy capacity only in vehicle communications. Under the assumption of ICT implementation, it is clear that vehicle communications will be regulated by law, and all vehicles will maintain legal speed limits within traffic regulations. In addition, when fully autonomous driving is realized, it is assumed that the vehicles will observe safety distances in order to avoid collisions with other objects or vehicles.

*A. Safety Distance*

Adaptive cruise control (ACC) is an important function in autonomous driving automation that regulates the speed and distance between at least two vehicles. ACC systems have to deal with collision avoidance in various situations. Consider, for example, three scenarios: 1) the "stop and go" scenario, 2) the "emergency braking" scenario, and 3) the "cut-in" scenario [11]. All of these scenarios include the notion of braking distance between one vehicle and a following vehicle. Generally, braking distance refers to the length that it takes a vehicle to come to a complete stop after the initial braking point. The braking distance includes the safety distance between a host vehicle and a target vehicle [12]. That is, at least the safety distance is longer than the braking distance. Autonomous vehicles may perform braking process. The theoretical braking distance $d_L$ is calculated as follows [13]:

$$d_L = v_0\left(t_a + t_b + \frac{t_c}{2}\right) + \frac{v_0^2}{(2a_{max})}, \quad (1)$$

where $v_0$ is the initial velocity of the host vehicle, $t_a$ is the response period, $t_b$ is the braking clearance period, $t_c$ is the breaking force application period, and $a_{max}$ is the maximum deceleration rate while braking.

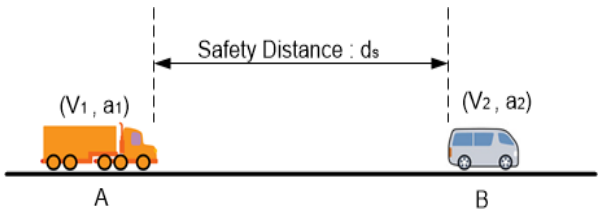

Figure 1. Safety distance for preventing an end collision

In Fig. 1, the braking safety distance shown is designed to prevent the host vehicle A from colliding with another vehicle, B. As shown in Fig. 1, when the host vehicle A detects that a collision is probable between the two vehicles (A, B), it automatically operates a safeguard in its intelligent cruise control system. Now, assume that the host vehicle A has the initial velocity $V_1$ and acceleration $a_1$, the target vehicle B has the initial velocity $V_2$ and acceleration $a_2$, and the safety distance is ds. Then, the safety distance $d_s$ is calculated as follows [14]:

$$d_s = v_1\tau + \frac{V_1^2}{2a_1} - \frac{V_2^2}{2a_2}, \quad (2)$$

where $\tau$ is a constant parameter of an ACC system mounted on the autonomous vehicle.

Our main concern of this paper is secure V2V communication. Owing to this, we assumed that host vehicle A and target vehicle B are in constant motion, at the same speed, based on their ACC systems. The respective ACC systems of host vehicle A and target vehicle B operate to maintain a certain distance between the first vehicle's front bumper and the second vehicle's rear bumper. Then, $V_1 = V_2$ and $a_1 = a_2$. As a result, the safety distance based on the host vehicle's velocity is represented as: $d_s = v_1\tau$. Consequently, we confirmed that the safety distance $d_s$ is proportional to the initial velocity $v_1$ of host vehicle A in V2V communication. That is, speed is proportional to safety distance in autonomous vehicles.

*B. Speed Limit*

All motorways have legally defined speed limits. When attempting autonomous driving, the concept of a speed limit is essential in that it will regulate the threshold for speed automatically. Speed limits, though, are applied differently on different roads. For example, in highway situations, the speed limit is relatively high. On the other hand, traffic is relatively slow in urban traffic conditions and thus the speed limit reflects this pattern. The speed limit, therefore, determines the minimum and maximum speeds a vehicle can travel for optimal traffic conditions and safety. For example, the highest highway speed may be 80 Km/h in the first direction, and 100 Km/h or higher in the second direction, while, in many countries, city road speed limits are typically set between 30 to 50 Km/h.

## C. Response Time and Collision Avoidance Time

One main factor associated with the safety distance is the response time in brake operation performance. For the case of partial autonomous driving, this response time significantly differs from the response that the user perceives and the actual tread. Even in the case of autonomous driving, this response time may vary depending on how the operation mode is set. In either case, the safety distance is known to affect the response time, as well as the fact that the safety distance is closely related to secrecy capacity [10]. Based on this, we can assume that the response time can be used to vary secrecy capacity. Collision avoidance time means the time until a vehicle accident is recognized, the collision is predicted, and the predicted result is propagated in V2V communications. The collision avoidance time may consist of the time required to transmit the message, the room wait time, and the time to analyze and calculate the likelihood of a defined collision analysis time collision [15].

## D. Secrecy Capacity

In information theory, Shannon channel capacity is known as a maximal amount of information that can be transmitted through a wireless channel. In general, channel capacity is given as
$$C = W\,log(1 + SNR), \qquad (3)$$
where W is the channel bandwidth, and SNR is the signal-to-noise ratio. Secrecy capacity denotes the channel capacity of a legitimate channel less the channel capacity of a wiretap channel. That is, secrecy capacity is a maximum data rate that is achievable between the legitimate TX-RX pair, subject to the constraints on information attainable by an unauthorized receiver [16]. For a Gaussian wiretap channel, secrecy capacity $C_s$ is:
$$C_s = \frac{1}{2} log\left(1 + \frac{P}{N_m}\right) - \frac{1}{2} log\left(1 + \frac{P}{N_w}\right), \qquad (4)$$
where P is the transmitter's power, $N_m$ is the receiver's noise, $N_w$ is the eavesdropper's noise.

## E. Physical Layer Security of V2V Communications

Physical layer security can be considered as a concept of allocating any one of a plurality of wireless channels to be radiated. That is, physical layer security will make very few radio signals to be transmitted to the eavesdroppers, or signals transmitted to the eavesdroppers will transmit signals that are completely different from those to which the legitimate sender is sent. As a result, this physical layer security ensures that a unique wireless channel is allocated between the sender and the receiver. Physical layer security can be considered as a concept of encapsulating any one of a plurality of radiating wireless channels, referring to Fig. 2. That is, according to physical layer security, there is almost no radio signal transmitted to the eavesdropper, or a signal completely different from the signal transmitted to the eavesdropper is transmitted to the legitimate sender. As a result, it can be appreciated that the role of such physical layer security is to allocate a unique wireless channel between the sender and the receiver.

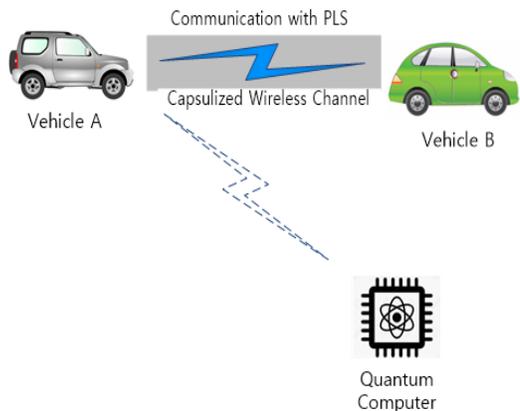

Figure 2. Secure V2V communication with PLS

When using physical layer security, the host vehicle can securely communicate with the target vehicle through the encapsulated wireless channel. So the threats of eavesdropping/tampering of quantum computers are completely gone. How can we ensure that wireless channels are securely cached by using physical layer security? The answer can be found in secrecy capacity that started from information theory.
Unlike traditional wireless communication channels, are there unique parameters that affect secrecy capacity when used only in V2V communications? When fully autonomous driving is realized, we assume that the vehicles will observe safety distances to avoid collisions with other objects or vehicles. There is a close relationship between the vehicle speed and secrecy capacity with respect to vehicle communications [10]. Based on the assumptions of the Information and Communication Technologies (ICT) implementation, it is clear that vehicle communications will be regulated by law and that all vehicles will maintain legal speed limits as per traffic regulations. The safety distance and speed limit are two parameters only used in vehicle communication, such that we must provide a detailed description of secrecy capacity and the relationship between it and vehicle communication. In general, the relationship between secrecy capacity according to the antenna-related parameters, beamforming, jamming, the size of the antenna and the number of antennas was also discussed. Ahn and Lee studied vehicle-related parameters that affect secrecy capacity of vehicle communication, such as vehicle speed, response time and speed limit. When considering Doppler effect according to the speed of the vehicle, secrecy capacity is studied to be proportional to the speed, but when considering the safety distance of autonomous driving, secrecy capacity is inversely proportional to the vehicle speed [17].

III. SYSTEM MODELING FOR V2V COMMUNICATIONS

Part of our objective for this study was to identify and locate elements that have limitations affecting secrecy capacity in the specific realm of vehicle communications. The speed limits and safety distances mentioned above were anticipated as factors that would affect secrecy capacity, and we studied their relationship in various scenarios. In particular, we assumed a highway situation environment, which is affected by safety distance and city road conditions, which are both, in turn, influenced by the speed limit, and proposed the corresponding system modeling.

*A. Highway System Model*

We expect that the distance between host vehicle A and target vehicle B is longer than safety distance Ds in a highway system. We assume that eavesdropper E is a significant distance away from host vehicle A and target vehicle B. That is, the distance r between host vehicle A and eavesdropper E is similar to distance r' between target vehicle B and eavesdropper E. Distance D between host vehicle A and target vehicle B is rθ, where θ is an angle that is formed by the first and second AE lines in Fig. 3.

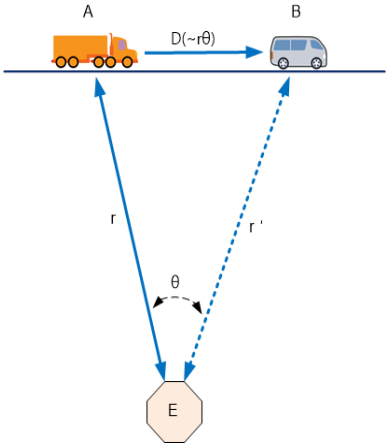

Figure 3. Highway system model

We consider V2V communication scenario for eavesdropper E. Secrecy capacity $C_s$ in the fading scenario is given as [18-21]:
$$C_s = \log_2\left(1 + \frac{P|h_{AB}|^2}{N_0}\right) - \log_2\left(1 + \frac{P|h_{AE}|^2}{N_0}\right), \quad (5)$$
where P is the transmission power of host vehicle A, $h_{AB}$ is a fading channel coefficient between host vehicle A and target vehicle B, $h_{AE}$ is a fading channel coefficient between host vehicle A and eavesdropper E, and $N_0$ is the variance of additive white Gaussian noise (AWGN).

In general, there are three fading models: 1) Rayleigh model, 2) Rician model, and 3) Nakagami model. The transmitted signal power may decrease with the distance as $d^{-\alpha}$, where α is the path loss exponent [22]. When the path loss distance reaches distance D(=rθ) between host vehicle A and target vehicle B, the fading channel coefficient $h_{AB}$ is given as : $h_{AB} = \left|\frac{1}{(r\theta)^\alpha}\right|$. In addition, when the path loss distance reaches distance r between host vehicle A and eavesdropper E, fading channel coefficient $h_{AE}$ is given as $h_{AE} = \left|\frac{1}{r^\alpha}\right|$.

Accordingly, in this system, the model secrecy capacity $C_s$ is:
$$C_s = \log_2\left(1 + \frac{P}{N_0(r\theta)^{2\alpha}}\right) - \log_2\left(1 + \frac{P}{N_0 r^{2\alpha}}\right). \quad (6)$$

We found that the velocity of the host vehicle is associated with safety distance ds between the legitimate terminals as $d_s = v_1\tau$. From the velocity point of view, distance D between host vehicle A and target vehicle B becomes $D = r\theta = v\tau$, where v is the current velocity of host vehicle A and $\tau$ is a constant parameter of the vehicle ACC system.

As a result, in this system model, secrecy capacity $C_s$ is:
$$C_s = \log_2\left(1 + \frac{P}{N_0(v\tau)^{2\alpha}}\right) - \log_2\left(1 + \frac{P}{N_0 r^{2\alpha}}\right). \quad (7)$$

For the simplicity of analysis, we assumed that distance r is fixed. Then secrecy capacity $C_s$ is a function that has only two variable parameters, v and α. Fig. 4 shows secrecy capacity $C_s$ versus vehicle velocity v.

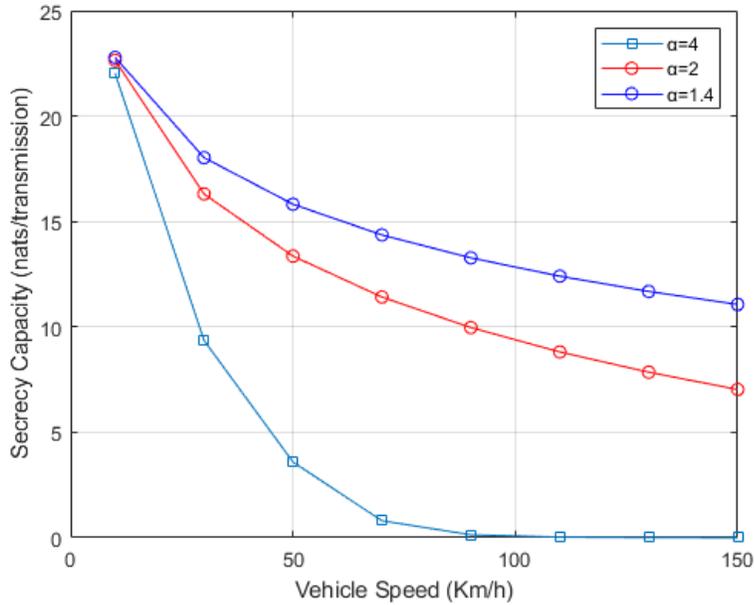

Figure 4. Secrecy capacity vs. vehicle speed for different path loss coefficients (α = 4, α = 2, and α = 1.4), for r = 1,000 m, P/N$_0$ = 70 dB, and τ = 200 ms

As shown in Fig. 4, secrecy capacity C$_s$ is strongly affected by path loss coefficient α and vehicle velocity v. First, when vehicle velocity v increases, secrecy capacity C$_s$ decreases, regardless of path loss coefficient α. Secondly, the greater the path loss coefficient α is, the greater secrecy capacity C$_s$ becomes. In addition, we expect that increasing the velocity of host vehicle A will decrease secrecy capacity C$_s$. For convenience of explanation, we assume that autonomous vehicles travel on expressways. For a simulation, Fig. 5 shows secrecy capacity C$_s$ for different velocities: 80 Km/h, 100 Km/h, and 120 Km/h. In this simulation, we assumed that distance r between A and E is 1000 m, and the fading channel model is Rayleigh model. As shown in Fig. 5, secrecy capacity at 80 Km/h is the highest, while secrecy capacity at 120 Km/h is the lowest. We confirmed that a vehicle traveling at a high speed may lose its secrecy capacity. Considering security environments, we need to optimize transmission power P or the transmission rate according to vehicle speed.

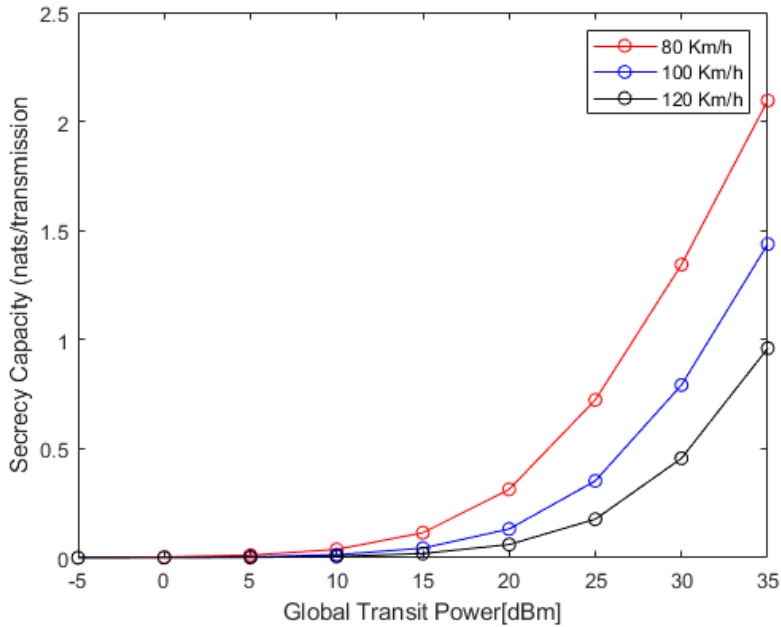

Figure 5. Secrecy capacity of different vehicle speeds (80 Km/h, 100 Km/h, and 120 Km/h) for path loss coefficient α = 3.5, r = 1000 m, and θ ≒ 0.1

As shown in Fig. 6, it is possible to find variations in secrecy capacity with transmission power ratio $P/N_0$. It can be seen that secrecy capacity is proportional to the magnitude of transmission power ratio $P/N_0$. That is that the greater the transmission power ratio $P/N_0$ is, the greater secrecy capacity $C_s$ becomes.

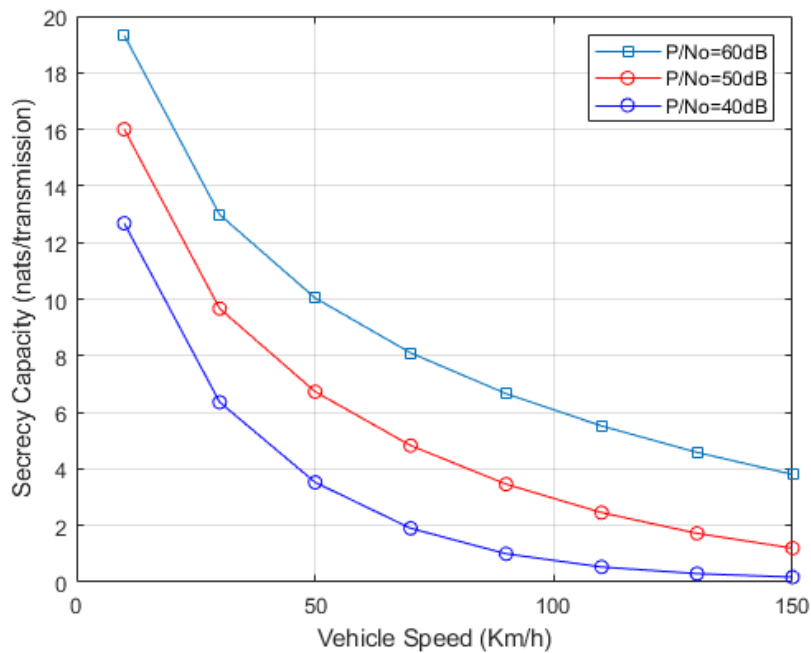

Figure 6. Secrecy capacity for different vehicle speeds and transmission powers ($P/N_0 = 40$ dB, $P/N_0 = 50$ dB, and $P/N_0 = 60$ dB) with $\alpha = 1.4$, $r = 1000$ m, and $\tau = 400$ ms

In addition, we examined how dependent the system secrecy capacity is on the reaction speed of ACC systems. As shown in Fig. 7, the faster the response speed of the system, the greater secrecy capacity.

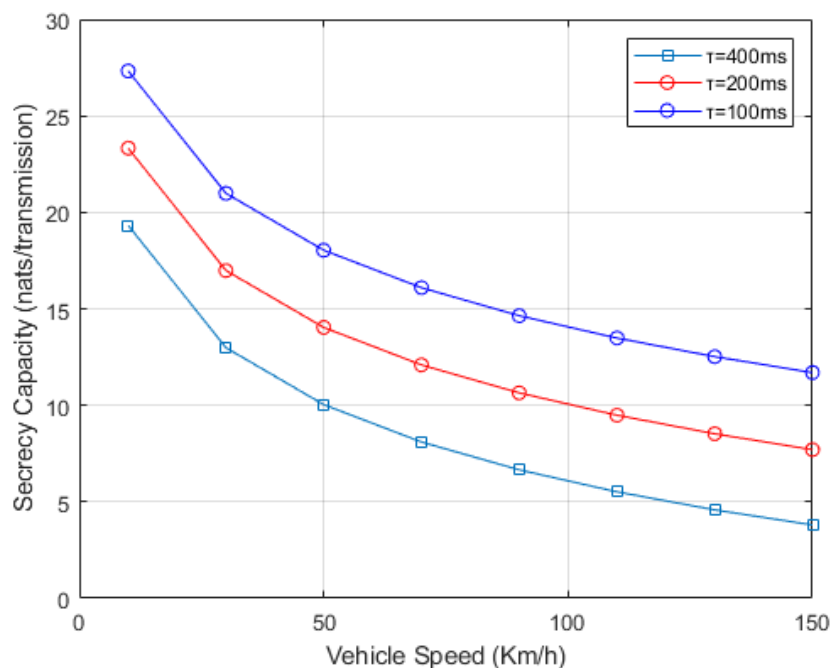

Figure 7. Secrecy capacity vs. vehicle speed for different response speeds ($\tau = 100$ ms, $\tau = 200$ ms, $\tau = 400$ ms) with $\alpha = 1.4$ and $r = 1000$ m

Accordingly, we have found that it is important to select the appropriate vehicle velocity and transmission power ratio $P/N_0$ to maintain a constant secrecy capacity. This makes it possible to define secrecy capacity as a major criterion for maintaining a certain level of security in vehicular communications. As described above, secrecy capacity of a vehicle can be determined by vehicle speed, magnitude of the transmission power, and system parameters. In summary, the secrecy capacity is inversely proportional to the speed of the vehicle, inversely proportional to the system parameters, ie response time, and proportional to the transmit power. However, this research needs to be extended, validated with theoretical analysis, and consider the speed of the eavesdropper as well [23].

## B. Urban traffic System Model

Vehicle communication modeling under real city road conditions is different from highway conditions. First of all, there are more vehicles in narrower spaces on city roads, and vehicle speed is relatively slower than when traveling on a highway. Here, it is also more probable that various kinds of eavesdroppers are present than on highways, as an eavesdropper can secretly collect vehicle communication information of slowly passing vehicles from a fixed location. Alternatively, an eavesdropper in a moving vehicle may steal vehicle communication information from another vehicle traveling in the vicinity completely undetected. On city roads, it is assumed that autonomous driving would be employed to maintain the speed limit. It can be assumed that the majority of autonomous vehicles would travel at the speed corresponding to the city speed limit that typically fall between 30 to 50 Km/h in many countries. It follows, then, that the speed of each vehicle traveling on any road is essentially fixed by the speed limit. Generally, when an eavesdropper targets vehicles on city roads, it is during communication attempts at intersections, or other critical moments while driving. This is because the most information is likely to be transmitted at intersections, for example the cross load. Of course, if eavesdropper E knows this, they would want to tap into the vehicle communication system when the vehicle is at such a fixed or stationary location. For the purposes of this study, we calculated secrecy capacity based on intersection speed limits. Our assumption is that both host vehicle A and target vehicle B are moving in directions perpendicular to each other while approaching the intersection, and that the speeds of host vehicle A and target B are both represented as $V_L$, referring to Fig. 8.

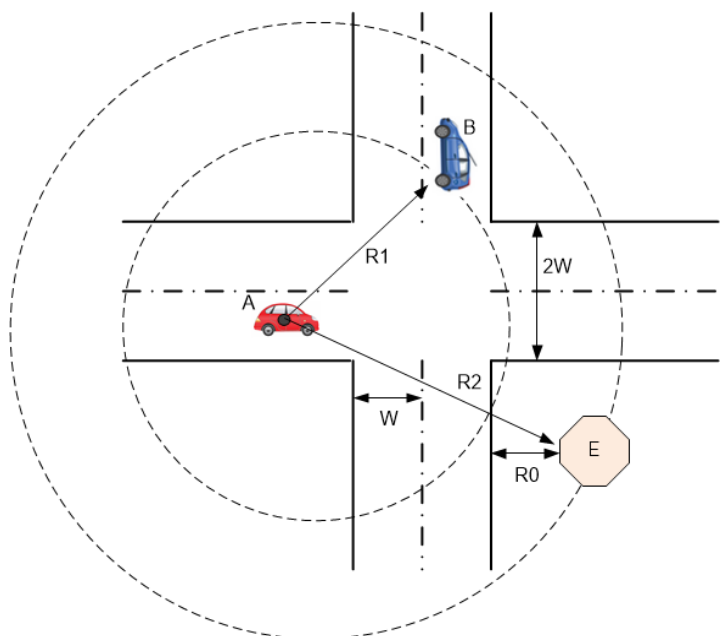

Figure 8. Urban traffic system model with a fixed eavesdropper

The moving distance of target vehicle B can be expressed by the following equation according to time, $S_T = 2W + V_L t$, where W is the width of the road line and $V_L$ is the speed limit.

For convenience of explanation, as shown in Fig. 9, if the host vehicle is at the southwest corner of the intersection, the travel distance of host vehicle A can be expressed by the following equation according to time, $S_H = W - V_L t$.

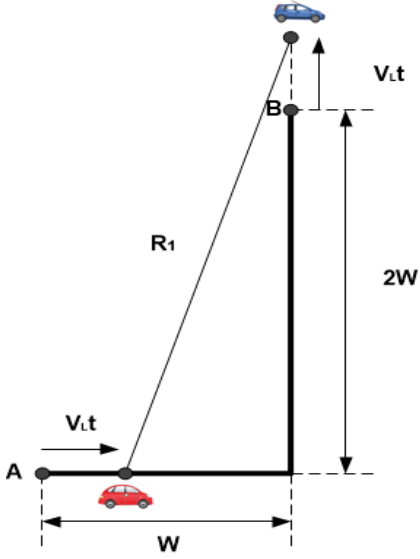

Figure 9. $R_1$ between host vehicle A and target vehicle B

Therefore, as shown in Fig. 10, distance $R_1$ between host vehicle A and target vehicle B can be expressed by the following equation based on Pythagorean Theorem:

$$R_1 = \sqrt{S_T^2 + S_H^2} = \sqrt{5W^2 + 2WV_L t + 2V_L^2 t^2}. \tag{8}$$

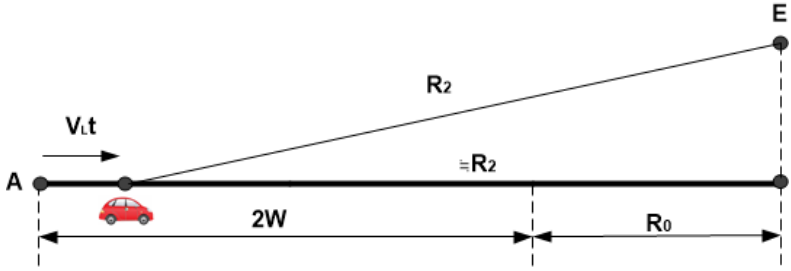

Figure 10. $R_2$ between host vehicle A and eavesdropper E

We assumed that distance $R_2$ between host vehicle A and eavesdropper E is significantly far away from $R_1$ in order to simplify the model. Under this assumption, distance $R_2$ between host vehicle A and eavesdropper E can be expressed as, $R_2 = R_0 + 2W - V_L t$, where $R_0$ is the shortest distance from the intersection to fixed eavesdropper E. Accordingly, in this system model, secrecy capacity $C_s$ is denoted:

$$C_s = log_2\left(1 + \frac{P}{N_0 R_1^{2\alpha}}\right) - log_2\left(1 + \frac{P}{N_0 R_2^{2\alpha}}\right) = log_2\left(1 + \frac{P}{N_0(5W^2 + 2WV_L t + 2V_L^2 t^2)^\alpha}\right) - log_2\left(1 + \frac{P}{N_0(R_0 + 2W - V_L t)^{2\alpha}}\right). \tag{9}$$

For the simplicity of analysis, we assumed that W=3m, P/N$_0$=70dB, t=0.1s, R$_0$=200m. Then secrecy capacity $C_s$ is a function expressed by only two variable parameters, $V_L$ and α. Fig. 11 shows secrecy capacity $C_s$ according to vehicle speed $V_L$.

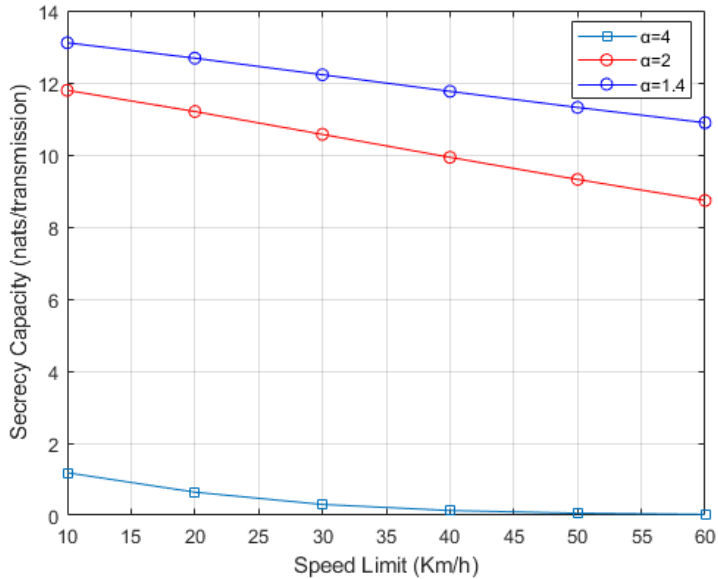

Figure 11. Secrecy capacity $C_s$ according to speed limit

The vehicle speed limit changed from 10 Km/h to 60 Km/h, and secrecy capacity was examined. As expected, secrecy capacity decreased as vehicle speed—and, therefore, speed limit—increased. When changing $R_0 = 20$ m under the same conditions in Fig. 11, the simulation result demonstrated that the overall secrecy capacity decreases with the speed limit. Thus, it was confirmed that the position of the fixed eavesdropper significantly affects secrecy capacity. In addition, we can see that the possibility of security failure is high because secrecy capacity has a minus value even under a certain speed. It would be reasonable to conclude that safe vehicle communication is not easily attained at specific ranges as indicated in Fig. 12.

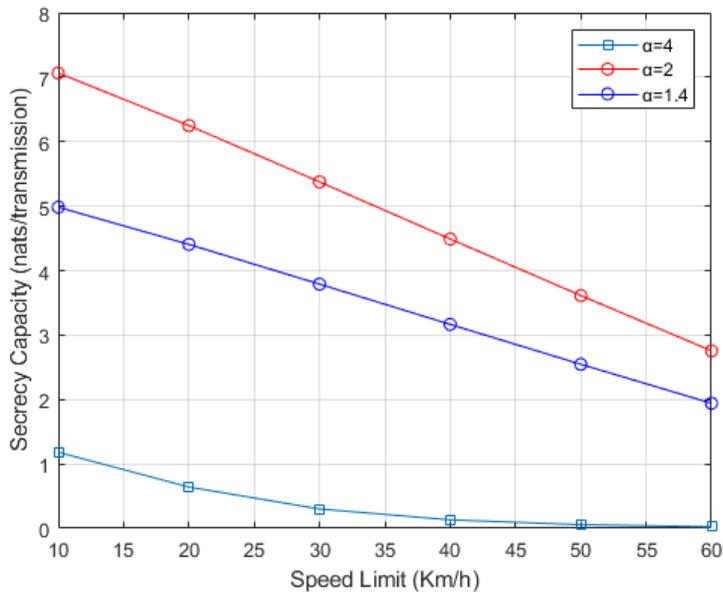

Figure 12. Secrecy capacity $C_s$ according to speed limit with $R_0 = 20$ m

It can be considered that increasing the transmission power may improve secrecy capacity. As such, we simulated transmission power by changing it from 70 dB to 80 dB. The results are shown the following Fig. 13. However, as seen in Fig. 13, there is no significant improvement in secrecy capacity even when transmission power is increased.

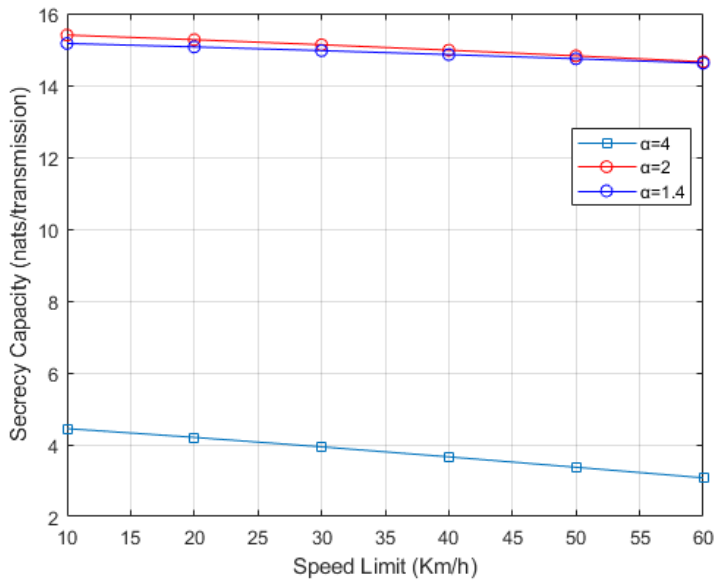

Figure 13. Secrecy capacity with transmission power 80dB

According to secrecy capacity based on a speed limit range, it can be seen that the position of the eavesdropper is very important, and it is relatively difficult to maintain secure secrecy capacity in vehicle communications on city roads compared to highway environments. Through the crossroads situation, we learned of the need to improve secrecy capacity through transmission power, vehicle speed, and other types of parameters. Below we consider secrecy capacity when the eavesdropper moves. The eavesdropper may think that similar speeds are moving within the cluster. When there is an eavesdropper in the same cluster, the system can be illustrated as follows. In particular, we assume an eavesdropper is running straight in the same direction as the target vehicle. This is a meaningful assumption if the attacker's vehicle is not just a passing member in urban traffic, but an active eavesdropper. As shown in Fig. 14, the diagram corresponding to the model is provided below.

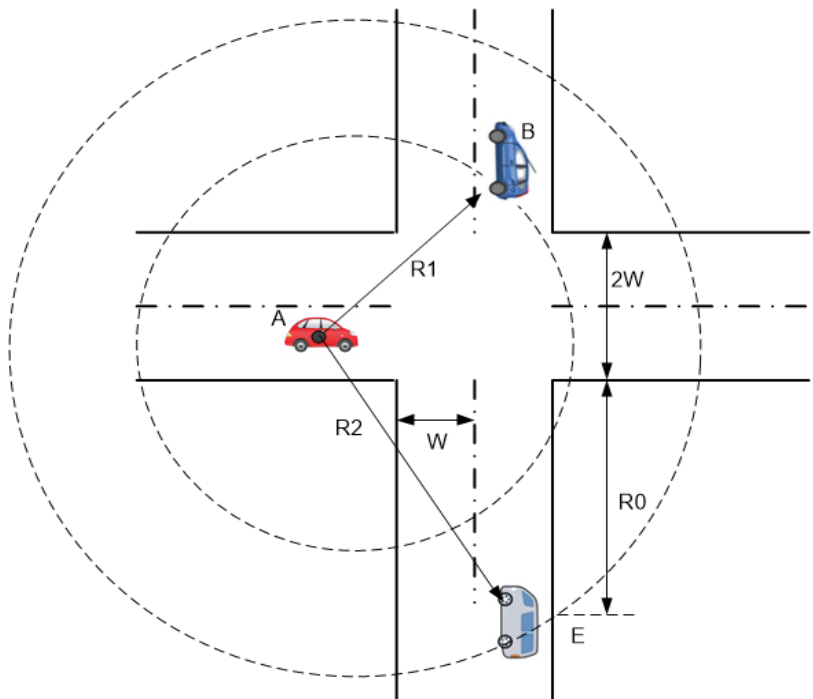

Figure 14. Urban traffic system model with a moving eavesdropper

For convenience in this study, it is assumed that both host vehicle A, target vehicle B, and eavesdropper E are traveling at the same speed. For instance, this could mean a limited speed encountered by in an autonomous driving situation. Therefore, distance $R_1$ between host vehicle A and target vehicle B can be expressed based on the Pythagorean Theorem:

$$R_1 = \sqrt{5W^2 + 2WV_L t + 2V_L^2 t^2}. \quad (10)$$

As such, distance $R_2$ between host vehicle A and eavesdropper E can also be expressed by the following equation based on the Pythagorean Theorem:

$$R_2 = \sqrt{W^2 + R_0^2 - 2(W + R_0)V_L t + 2V_L^2 t^2}. \quad (11)$$

Accordingly, in this system model, secrecy capacity $C_s$ is:

$$C_s = log_2\left(1 + \frac{P}{N_0 R_1^{2\alpha}}\right) - log_2\left(1 + \frac{P}{N_0 R_2^{2\alpha}}\right) =$$
$$log_2\left(1 + \frac{P}{N_0(5W^2 + 2WV_L t + 2V_L^2 t^2)^\alpha}\right) - log_2\left(1 + \frac{P}{N_0(W^2 + R_0^2 - 2(W + R_0)V_L t + 2V_L^2 t^2)^\alpha}\right). \quad (12)$$

For simplicity of analysis, we assumed that W=3m, $P/N_0$=70dB, t=0.1s, $R_0$=20 m. If this is true, then secrecy capacity $C_s$ is a function expressed only by two variable parameters, $V_L$ and α. Fig. 15 shows secrecy capacity $C_s$ according to vehicle speed $V_L$. The vehicle speed limit changed from 10 Km/h to 60 Km/h, and secrecy capacity was examined. As expected, secrecy capacity decreased as the speed limit increased.

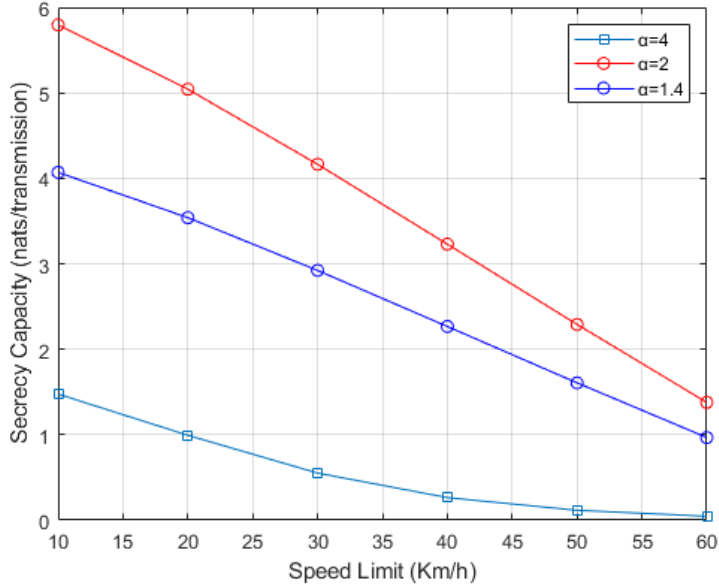

Figure 15. Secrecy capacity $C_s$ according to speed limit against a moving eavesdropper

## C. Relay-based System Model

As described above, vehicle secrecy capacity decreases as the vehicle speed increases. Decreased secrecy capacity with increasing vehicle speed can be compensated by cooperative relay communication. In V2V communication, secrecy capacity can be improved by adopting one relay R between host vehicle A and target vehicle B. In general, secrecy capacity of cooperative relay communication is higher than that of direct communication without a relay [24]. For simplicity of the analysis of secrecy capacity, we assume that the system model comprises one relay R between the host vehicle A and the target vehicle B, as shown in Fig. 16.

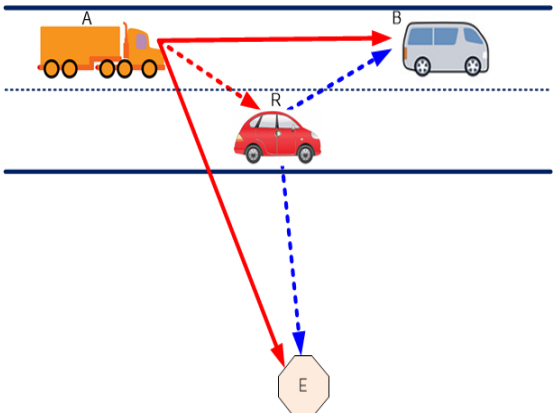

Figure 16. Relay-based System Modeling

The channel capacity of the legitimate channel is expressed as:

$$C_1(A,B) = W \log_2\left(1 + \left(\frac{P_A h_{AB}}{P_R h_{RB} + \sigma_B^2}\right)\right), \quad (13)$$

where $P_A$ and $P_R$ are, respectively, transmission powers of host vehicle A and relay R, $h_{AB}$ is channel gain between host vehicle A and target vehicle B, $h_{RB}$ is channel gain between relay R, and target vehicle B, $\sigma_B^2$ is the AWGN in target vehicle B, and W is bandwidth. The channel capacity of the wiretap channel is shown as:

$$C_2(A,E) = W \log_2\left(1 + \left(\frac{P_A h_{AE}}{P_R h_{RE} + \sigma_E^2}\right)\right), \quad (14)$$

where $h_{AE}$ is channel gain between host vehicle A and eavesdropper E, and $\sigma_E^2$ is the AWGN in target vehicle B. Then, secrecy capacity with the cooperative relay communication is denoted:

$$C_R = W\left[\log_2\left(1 + \left(\frac{P_A h_{AB}}{P_R h_{RB} + \sigma_B^2}\right)\right) - \log_2\left(1 + \left(\frac{P_A h_{AE}}{P_R h_{RE} + \sigma_E^2}\right)\right)\right]. \quad (15)$$

Fig. 17 shows secrecy capacity with and without relay R. Referring to Fig. 17, relay R helps to improve the overall secrecy capacity. We confirmed that V2V communication using the relay may enhance secrecy capacity, and the relationship between the existence of the relay and secrecy capacity.

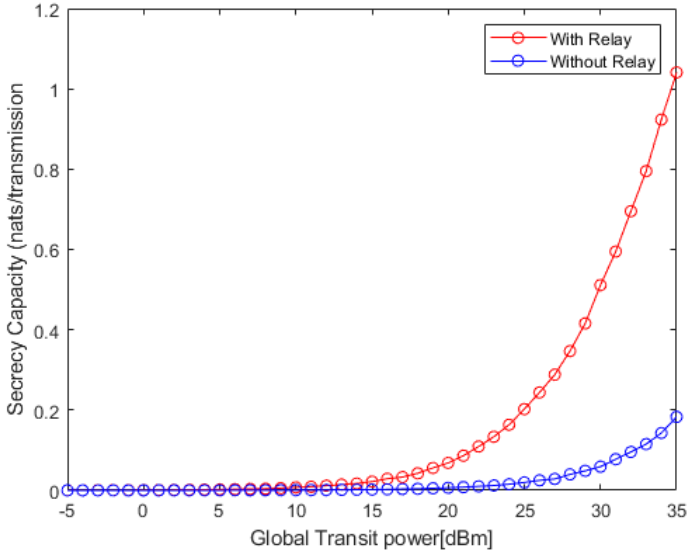

Figure 17. Relay dependence of secrecy capacity

## IV. VEHICLE SECRECY CAPACITY

We have looked at approximately how various factors of a vehicle may affect secrecy capacity. However, as the definition of secrecy capacity itself is based on the existence of a known eavesdropper, it is difficult to calculate secrecy capacity in a real environment. Therefore, in order to actually use secrecy capacity in vehicle communication, the definition of secrecy capacity must be determined in any way. We have examined the definition of secrecy capacity discussed so far and defined the concept in terms of real vehicle communication.

### A. Ergodic Secrecy Capacity

Generally speaking, secrecy capacity does not consider temporal variations in the communication channel. Thus, we introduce the concept of ergodic secrecy capacity. Ergodic secrecy capacity is defined as the time average of secrecy rate over a legitimate user and an eavesdropper [25-27]. For an ergodic fading channel in multiple-input and multiple-output (MIMO) systems, fading channel coefficients are independent and identically distributed. Thus, in the vehicular network, host vehicle A, target vehicle B, and eavesdropper E each may experience different fading respective channels. Assuming that all terminals have perfect channel state information (CSI) about the current fading state, ergodic secrecy capacity is defined by [28]:

$$C_S = \max_{E_A[\gamma] \leq P} \boldsymbol{E}_A \log_2\left(1 + \frac{\gamma|h_{AB}|^2}{\sigma_B^2}\right) - \log_2\left(1 + \frac{\gamma|h_{AE}|^2}{\sigma_E^2}\right), \quad (16)$$

with γ denoting power allocation and
$$A = \left\{h_{AB}, h_{AE} : \frac{|h_{AB}|^2}{\sigma_B^2} > \frac{|h_{AE}|^2}{\sigma_E^2}\right\}.$$

## B. Secrecy Capacity using the Poisson Point Process

In the real world, eavesdroppers are likely to be randomly located rather than stationary in some fixed position. Therefore, the more meaningful calculation of secrecy capacity needs to be modified. We address a model of secrecy capacity that considers eavesdroppers distributed according to Poisson point process (PPP). Ghogho and Swami also introduce a Poisson random field of eavesdroppers based on Rayleigh fading model and PPP [29]. In our model, we assumed that the positions of the eavesdroppers are unknown to the legitimate vehicles (A or B) and are randomly distributed according to PPP. The probability of finding n eavesdroppers is set out in the Poisson distribution:

$$f(n, \lambda) = \frac{\lambda^n e^{-\lambda}}{n!}, \tag{17}$$

where λ is the expected number of eavesdroppers in a predetermined area. Secrecy capacity of the above model is calculated by [30, 31]:

$$C_s = log_2(1 + SNR_{AB}) - log_2\left(1 + \pounds(SNR_{AE})\right) \cong log_2\left(\frac{|h_{AB}|^2 d_{AB}^{-\alpha}}{\pounds\left(|h_{AE}|^2 d_{AE}^{-\alpha}\right)}\right), \tag{18}$$

where $h_{AB}$ is the fading coefficient between vehicle A and vehicle B, $h_{AE}$ is the fading coefficient between vehicle A and eavesdropper E, $d_{AB}$ is the distance between vehicle A and vehicle B, $d_{AE}$ is the distance between vehicle A and eavesdropper E, α is the path loss exponent, $\pounds(\cdot)$ is either $\sum_{e\in\Phi}(\cdot)$ for the colluding scenario or $\max_{e\in\Phi}(\cdot)$ for the non-colluding scenario, and Φ denotes PPP of eavesdroppers. Assuming that there are N eavesdroppers, the average secrecy capacity can be calculated as:

$$C_{savg} = \frac{\sum_{i=1}^{N}\left(log_2(1+SNR_{AB}) - log_2\left(1+SNR_{AE_i}\right)\right)}{N}, \tag{19}$$

The largest secrecy capacity that corresponds to each of the eavesdroppers selected by PPP will be equal to or greater than the average secrecy capacity $C_{savg}$ ($C_{savg} \leq C_{max}$). Here, the largest secrecy capacity is determined by the closest eavesdropper to host vehicle A.

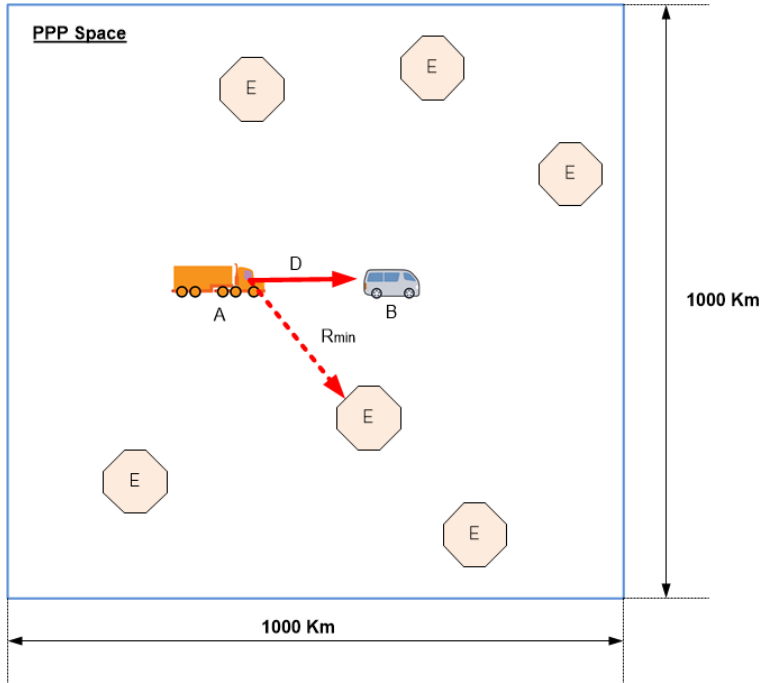

Figure 18. Simulated environment for PPP

We assumed that there are six eavesdroppers per 1,000 m² in Fig. 18. We attempted to calculate the average secrecy capacity based on the probability that six eavesdroppers would be in real time by PPP. The simulation results for secrecy capacity are shown in Fig. 18, which are based on the assumption that the average secrecy capacity is determined by one eavesdropper (among the six) who is the closest to host vehicle A. For the sake of explanation, we assumed that target distance D between host vehicle A and target vehicle B is shorter than minimum distance $R_{min}$. Here, $R_{min}$ is the distance from host vehicle A to the nearest eavesdropper.

For example, in Figs. 19A, 19B, and 19C, the target distances are 0.1 $R_{min}$, 0.3 $R_{min}$, and 0.5 $R_{min}$. As shown in Figs. 19A, 19B and 19C, we confirmed that the average secrecy capacity can be conceptually calculated in real time using PPP.

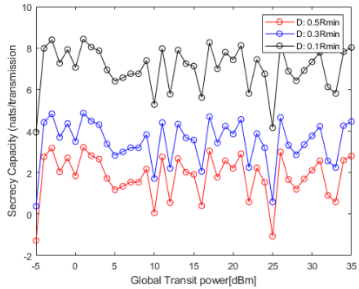 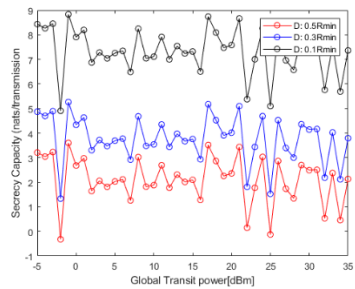 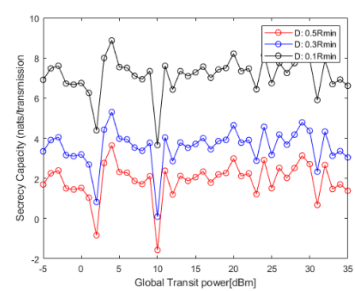

Figure 19A. Secrecy capacity at T1    Figure 19B. Secrecy capacity at T2    Figure 19C. Secrecy capacity at T3

*C. Vehicular Secrecy Capacity*

We have examined, approximately, how various factors of a vehicle may affect secrecy capacity. However, since the definition of secrecy capacity itself is based on the existence of a known eavesdropper, it is difficult to calculate secrecy capacity in a real environment. Therefore, to actually use secrecy capacity in vehicle communication, we must determine the definition of the term. Thus far, we have examined the definition of secrecy capacity based on our discussion and defined the concept in terms of real vehicle communication. In the real world, eavesdroppers are likely to be randomly located rather than stationary at one fixed position. Therefore, we must develop a more relevant secrecy capacity calculation method. We address a model of secrecy capacity that considers eavesdroppers distributed according to Poisson point process (PPP). Our secrecy capacity, referred to as VSC, is similar since it is independent of the existence of an eavesdropper. In general, the eavesdropper may or may not transmit response messages corresponding to its channel information after receiving a communication-initiated signal from the host vehicle. The response message includes channel information that is transmitted through channel states, which may contain SNR values. A host vehicle may then receive channel information from either eavesdroppers or legitimate vehicles. VSC value is defined based on the assumption that SNR value of the eavesdropper is lower than the average SNR value. Using SNR values of the received channel information, host vehicles can define VSC as follows:

$$VSC = log_2(1 + SNR_{AB}) - log_2(1 + SNR_{XOR}), \qquad (20)$$

where $SNR_{XOR} = \frac{\sum_{i=1}^{M} SNR_{Ai}}{M}$, $M$ is the number of channel signaling data received during the unit time, $B$ represents the target vehicle for communication, and $i$ is the vehicle number, excluding the host vehicle, with reference to Fig. 20.

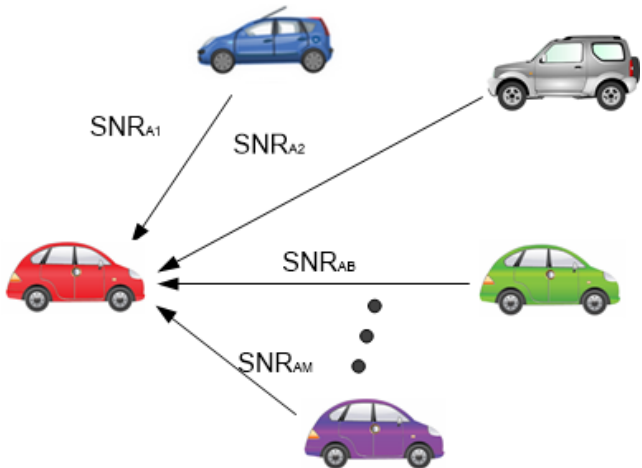

Figure 20. Definition of Vehicular Secrecy Capacity

Pinpointing an eavesdropper remains difficult. Due to this difficulty, we initiate VSC with an inferred eavesdropper SNR value, which is obtained using the mean SNR value in the channel information, including that of an eavesdropper. Although these VSCs are less accurate than the intended secrecy capacity, they should still produce critical information that determines the need for security and how changes affect the communications. For example, a potential or actual eavesdropper may determine that the vehicle communication environment is inferior when VSC value is below the reference value. In this case, we can determine that the security through the physical layer is limited and that we should employ additional security methods in vehicle communication. If VSC value is consistently low every week, vehicle communication may be inhibited for a predetermined period of time.

## V. PROPOSED V2V COMMUNICATION IN A SECURITY CLUSTER

The goal for this study is singular: to demonstrate a model where data transmission can be performed using secrecy capacity in future vehicle communications. Based on our knowledge, secrecy capacity is controllable in real-time wireless communications. Previous studies have already proposed vehicle-to-vehicle communication by maintaining a secrecy capacity constant [10]. Our model also allows vehicles with a certain level of secrecy capacity to form a security cluster, which indicates that vehicles in a security cluster are free to communicate with each other. Security clusters can be defined in a variety of ways, where secrecy capacity is the primary criterion. Next, aside from secrecy capacity, the direction of vehicle movement and speed are also major factors. An RSU may form a secure cluster while only vehicles with a block chain may form a secure cluster by themselves. We conclude that further studies on security clusters using block chains are necessary. We suggest the formation of a security cluster with a block chain technique using inherent vehicle values, such as the vehicle identification number (VIN).

Recent studies have also attempted to calculate the vehicular communication capacity over 5G networks [32-35]. In connection with this, and as described above, we confirmed the relationship between vehicle speed and secrecy capacity, as well as the relationship between the existence of a relay and secrecy capacity. We propose a new V2V communication system that uses cooperative relay communication based on secrecy capacity, as shown in Fig. 21.

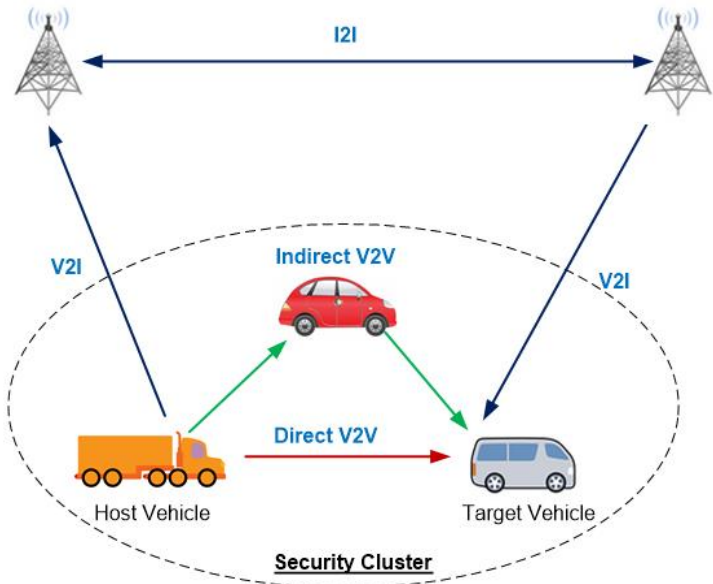

Figure 21. The proposed V2V communication system

### A. SC-based Communication

We propose a new V2V communication system that uses cooperative relay communication based on secrecy capacity, as shown in Fig. 22.

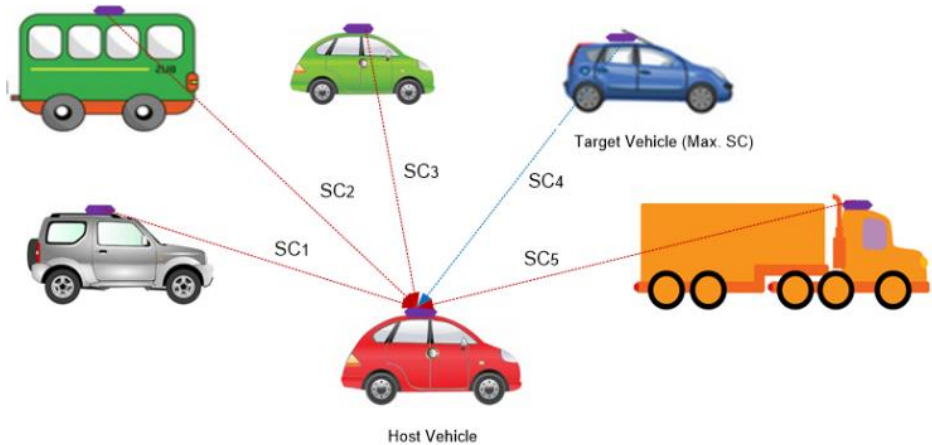

Figure 22. The SC-based V2V communication system

SC-based V2V communication is the communication method for the most basic secrecy capacity. The host vehicle receives channel information from each vehicle, calculates a secrecy capacity for the vehicle based on the received information, selects the target vehicle using the calculated secrecy capacity of the vehicle, and performs communication with the selected target vehicle. Here, the criterion for target vehicle selection can be determined to select the vehicle whose secrecy capacity has the largest value. On the other hand, target vehicle selection can be implemented to define a reference value for secrecy capacity and perform vehicle communication accordingly.

| SC-based Vehicle Communication Algorithm |
| --- |
| Step 1: Receive CSI information from vehicles. |
| Step 2: Calculate vehicular secrecy capacity related to each of the vehicles. |
| Step 3: Select a target vehicle among the vehicles using the calculated vehicular secrecy capacity. |
| Step 4: Communication with the selected target vehicle. |

*B.* RSC-based Communication

The reference secrecy capacity (RSC) can be defined as a value that determines whether or not communication occurs.

| RSC-based Vehicle Communication Algorithm |
| --- |
| Step 1: Receive CSI information from vehicles. |
| Step 2: Calculate vehicular secrecy capacity related to each of the vehicles. |
| Step 3: Select a target vehicle among the vehicles using the calculated vehicular secrecy capacity. |
| Step 4: Determine that VSC(Target) ≥ RSC. |
| Step 5: If VSC(Target) ≥ RSC, communicate with the selected target vehicle. |
| Step 6: Else, increase VSC(Target) using vehicle secrecy parameters and repeat step 5. |

Vehicle communication using RSC is for communication only when secrecy capacity of the target vehicle is more than a certain value, referring to Fig. 23. If RSC cannot be secured, secrecy capacity will be increased using at least one of the vehicle secrecy capacity parameters mentioned. This process can be repeated until the vehicle attains a certain level of secrecy capacity.

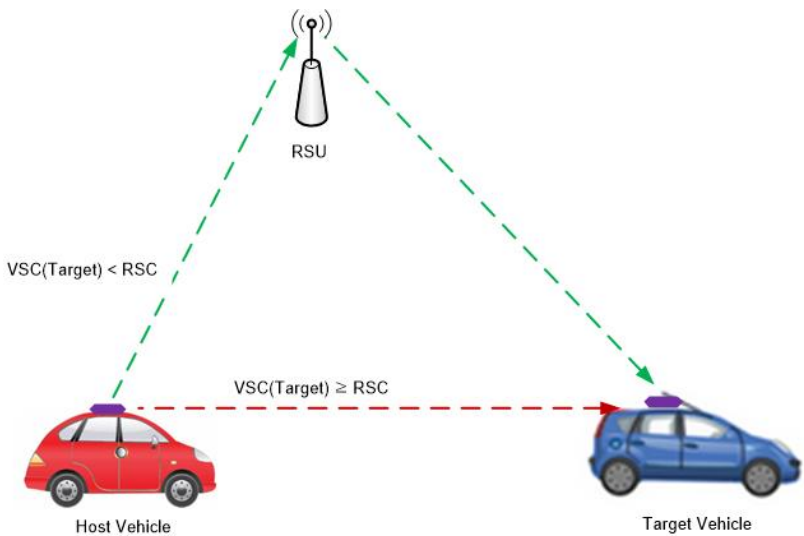

Figure 23. RSC-based V2V communication system

If secrecy capacity does not exceed the predetermined value, the host vehicle may initiate indirect communication with the target vehicle, which includes cooperative relay communication. Relay selection of the vehicular network is performed using an optimized algorithm that considers power consumption and secrecy enhancement.
The optimized algorithm, for example, can control the transmission power of the host vehicle to increase security in V2V communication system. V2V communication system enhances physical layer security due to cooperative relay communication based on secrecy capacity. On the other hand, if secrecy capacity does not exceed the predetermined value, the host vehicle may increase its transmission power by a predetermined amount. Thereafter initiating vehicular communication.

Meanwhile, the host vehicle can select the V2I communication, such as a 5G mobile network, according to the results of secrecy capacity calculation. In addition, we define a new CSI for 5G vehicular communications. The main content that we propose is included in SNR value of CSI. When attempting to perform initial communication, we can calculate SNR value at the receiving end if the receiver knows the power to be transmitted because the receiver is aware of the received power. All receivers can calculate SNR values and include them in their CSI. For a viable proposal, we must first generate sufficient discussion in the relevant standards associations. In general, CSI includes the channel quality information (CQI). We recommend that 5G communication includes an index of SNR as the CQI. Eavesdroppers should not know CSI. Our concern is with calculating secrecy capacity in this situation. However, most vehicles are equipped with radar, which can interact with wireless communication devices. Accordingly, the user can specify a wireless communication device that does not provide CSI. This will allow us to approximately predict CSI for eavesdroppers, even if the eavesdroppers do not provide a CSI. On the other hand, as mentioned above, the reference secrecy capacity may be different depending on the speed of the vehicle. In general, for a high-speed road, the reference secrecy capacity can be set to a small value as compared with that of a city traffic situation. If a certain amount of secrecy capacity cannot be secured, V2V communication can be abandoned and indirect V2V communication can be performed. We also aim to introduce the concept of security clusters into vehicle communications. Security clusters can be thought of as words but can also be viewed as a collection of all terminals that retain secrecy capacity at a certain level. A set of vehicles can be designated as security clusters and maintained as such.

*C.* SC Cluster-based Communication

The concept of a secure cluster is simply a group with a certain level of secrecy capacity. The goal of this study is singular: to demonstrate a model where data transmission is performed using secrecy capacity in future vehicle communications. Based on our knowledge, secrecy capacity can be controlled via real-time wireless communication. Previous studies have already proposed vehicle-to-vehicle communication by maintaining a constant secrecy capacity value [10]. Our model also allows vehicles with a certain level of secrecy capacity to form a security cluster, where vehicles in the security cluster can freely communicate with each other. Security clusters can be defined in a variety of ways, with a secrecy capacity as the primary criterion. Next, aside from secrecy capacity, the direction of vehicle movement and speed are major factors. An RSU may form a secure cluster, where only vehicles, such as a block chain, may form the secure cluster. We conclude that further studies on security clusters using block chains should be conducted. We recommend the formation of a security cluster with a block chain technique using inherent values of the vehicle, such as the vehicle identification number (VIN). Recent studies have also attempted to calculate the vehicular communication capacity over 5G mobile networks. As described above, we confirmed the relationship between vehicle speed and secrecy capacity, as well as the relationship between the existence of a relay and secrecy capacity. We propose a new V2V communication system that uses cooperative relay communication based on secrecy capacity.

The proposed SC Cluster V2V communication algorithm is as follows:

| SC Cluster-based Vehicle Communication Algorithm |
| --- |
| Step 1: Search target vehicle candidates using the vehicular secrecy capacity.<br>Step 2: Select valid vehicles among the target vehicle candidates.<br>Step 3: Configure security cluster using valid vehicles.<br>Step 4: Communication in the security cluster. |

The host vehicle may preferentially form a security cluster to communicate with the target vehicle. The process of forming a security cluster is as follows. The host vehicle can search the received signals for those whose vehicle secrecy capacity is equal to or greater than the reference value. The vehicles above the reference value are the target vehicle candidates.

Targeted vehicle candidates are at risk of selecting a malicious communication subject if they are immediately used for vehicle communication. To mitigate this danger, we must restrict communications to only vehicles. For this purpose, we can use unique information that can confirm the identity of the vehicle. For example, such vehicle specific information may be its VIN. When transmitting the initial SNR value to the target candidate vehicle, the hash value corresponding to the VIN can be simultaneously transmitted to the host vehicle. A nonce chain consisting of consecutively hashed values from the VIN may be used. By verifying the VIN with the verified hash value, we can confirm that the target vehicle candidate is a valid vehicle. Therefore, valid vehicles can be selected among the target vehicle candidates. Then, a security cluster corresponding to the target vehicle may be formed with the valid vehicles. The host vehicle is able to communicate with any of the target vehicle candidates within the security cluster. Our proposed vehicle communication method can perform communication simply and securely only with the existing SNR values. This is applicable to practical communication, with a low cost for the construction safety communication devices.

*D.* Security Cluster Management

Each vehicle can store or manage the members of the security cluster that performed V2V communication. The vehicle stores previous security cluster members in the database. The vehicle can also store current security cluster members in the database. It is, of course, one of the purposes of managing the history of a Luster member. There may be circumstances in which a target vehicle candidate that is not currently included in the security cluster but is available for reference in carrying out vehicle communication may have to be selected. In this case, if there is a vehicle included in the member of the previous security cluster in the history database, the vehicle can be selected as the target vehicle to perform the vehicle communication. The host vehicle can form a security cluster in real time and use this cluster to communicate with the vehicle for a certain period of time. Here, the security cluster may be composed of vehicles having a secrecy capacity equal to or greater than a predetermined reference value, as described above. In the case of a vehicle in which secrecy capacity does not satisfy the reference value as needed, a pseudo cluster may be formed based on the secondary reference value. Such pseudo-clusters may be incorporated into a secure cluster at any time by controlling secrecy capacity parameters of the host vehicle. Security clusters can be difficult to form in real time. A security cluster may be formed using a deep learning technique.

Generally, when considering commuting vehicles in urban traffic conditions, the probability of the same vehicles operating at the same time in the same place is very high. When modeling through learning, a security cluster can be easily configured, and selection of a target vehicle and secure communication with the target vehicle can be performed. Chen et. al introduced an rear-end collision prediction using deep-learning [36]. The forward-backward collision prediction is based on the safety distance, so it can be easily converted to secure capacity. That is, members of a security cluster over a specific value using deep learning are predictable. A security cluster model for a vehicle whose secrecy capacity is greater than or equal to a certain value can be generated through learning by deep learning. In the real-time vehicle communication, the host vehicle can easily determine whether or not the security cluster is incorporated by using only the information of the security cluster model and the target vehicle candidates.

Recently, blockchain technology has been introduced for secure vehicle communication [37 - 42]. A blockchain is a very attractive technology for those who want to be kept secret without the central control. It is recognized as a very useful and popular technology in IoT based environment. Blockchain can be classified into PoW (Proof-of-Work), PoS (Proof-of-Strake), PoN (Proof-of-Nonce), etc. according to the consensus node selection method [43]. The consensus node may be determined as a vehicle that can prove that secrecy capacity is greater than or equal to a certain value.

Then, how do you verify in other vehicles that secrecy capacity is above a certain level? A host vehicle that desires to communicate or desires to form a secure cluster may send a request message to neighboring vehicles that it wishes to form a secure cluster. The neighboring vehicles receiving the request message can determine themselves as candidates of the consensus node with their secrecy capacity exceeding a predetermined value, and transmit the response message to the host vehicle. The host vehicle may select a consensus node from the received response message based on a predetermined algorithm and send a consensus node selection result message corresponding to the selection result to the corresponding vehicle.

If the consensus nodes are selected by the above-described method, a block chain for vehicle communication can be formed. That is, a block to be coupled to the block chain can be generated periodically or non-periodically by agreement between the cone sensor nodes. On the other hand, proof that secrecy capacity is more than a certain value may not be easy. A malicious vehicle may disclose information that is deceptive even if its secrecy capacity is above a certain value. In order to prevent such malicious behavior, the host vehicle must be able to analyze the signaling information obtained from the target vehicle and verify whether the information related to secrecy capacity of the target vehicle is correct. Wherein the signaling information may comprise various information related to the information received in the subject vehicle in response to the initiation signal transmitted from the host vehicle. For example, SNR value may be included in the signaling information.

On the other hand, is secrecy capacity actually fixed? We confirmed that this was not the case in our previous study [10]. Factors that inherently affect secrecy capacity during vehicle communication include vehicle speed, response time, and transmission power. If secrecy capacity of the target vehicle is less than the reference value, we can perform vehicle communication while varying these factors to maintain a certain level of secrecy capacity. Confirming that secrecy capacity is controllable using these parameters in vehicle communication is important. In general, to improve secrecy capacity in wireless communication, relays and artificial noise (jamming) should be analyzed. However, there is a considerable cost associated with the use of relays and artificial noise to increase secrecy capacity. There is a need for an efficient method that increases secrecy capacity at a low cost. In our previous study, we mention that this method may be a compressive sensing technique [10].

## VI. Physical Layer Security Enhancement

In the real world, secrecy capacity may not always be strong. We have conceptually examined factors that affect secrecy capacity, so we assume that each vehicle commands some degree of control over secrecy capacity. However, determination of physical layer security is lacking with secrecy capacity alone.

*A. Compressive Sensing*

We propose a method to increase physical layer security by using compressive sensing, which is a recent development. Physical layer security using compressive sensing is expected to be efficient in image data transmission. This area promises to open doors for a vast range of research topics in the future. Recently, compressive sensing has been proposed as a solution for improving physical layer security [10, 44, 45]. Compressive sensing can theoretically be performed with less sampling than Nyquist sampling because compressive sensing integrates sampling and compression. Compressive sensing is described as $\boldsymbol{y = Ax = A\Psi s = \Omega s,}$ where y is a measurement vector, A is a sensing matrix, x is a signal vector, $\Psi$ is an orthogonal basis matrix, $\Omega$ is a system matrix, and s is a sparse vector. If x is plain-text and A is a cryptographic key, y can be interpreted as a cipher-text, as can be seen in Fig. 24.

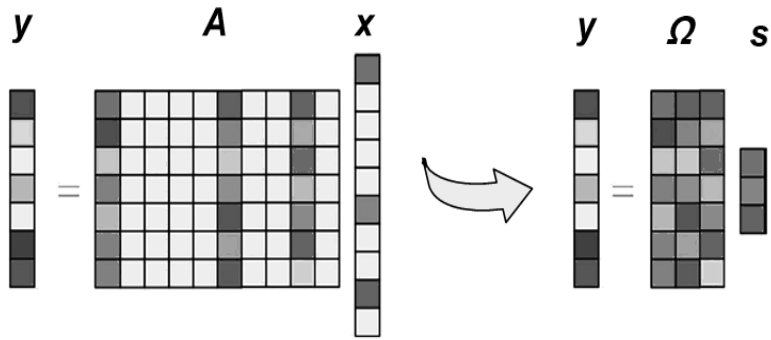

Figure 24. Compressive sensing concept

Basically, V2V communication satisfies physical layer security based on secrecy capacity. We can enforce security using physical layer encryption based on compressive sensing. Compressive sensing encryption can be an attractive solution as it can provide reasonably secure transmissions with simple low-complexity ciphers in the physical layer over wireless networks [46-51].

In the following description, a vehicular communication message (e.g., BSM) can be transmitted with compressive sensing encryption. Referring to Fig. 25, the first vehicle may compress BSM using compressive sensing encryption and output the compressed BSM externally via wireless communication. The second vehicle can then receive the compressed BSM via wireless communication for decoding.

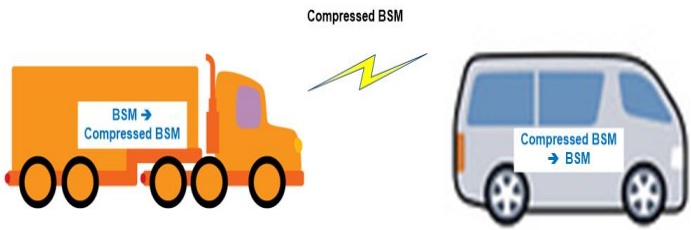

Figure 25. V2V communication with a compressed BSM

Further research into this topic will be needed in the future. If we can define BSM to satisfy the scarcity requirement, safe and inexpensive vehicular communication based on CS algorithm can be anticipated over 5G. This is because it would be virtually impossible to eavesdrop on such vehicular communication, and it does not require a complicated algorithm on the upper network layer-the application layer-to guarantee security. In the previous section, the physical layer security owing to compressive sensing was described in isolated terms for message transmission such as BSM. In the future, we intend to study the application of the acquired images in vehicles. When an image is to be transmitted using compressive sensing, it is expected that the amount of data transmitted will reduce along with communication costs. This feature has great potential to become an attractive element in vehicular communication.

*B. Offloading Service for Data Encryption*

Data encryption schemes are another area of research, as opposed to securing channels in physical layer security. In physical layer security, it is the protection of the data itself transmitted to the channel. This may not be different from traditional encryption techniques. Nonetheless, data encryption techniques, along with techniques that use compression sensing, play an important role in protecting data transmitted over a wireless channel. Data encryption is the most common method used to protect data information. In IEEE P1609.2, an elliptic curve encryption algorithm with a length of 256 bits provides the application-related VANET network and management of the proposed security mechanism. The proposed algorithm has a better effect than the RSA encryption algorithm and is considered to be a generic next generation public key cryptosystem. However, the signing and encryption of this encryption algorithm is fast and has the effect of time delay in large network applications. The problem is whether this data encryption technique can sufficiently exert its performance in an autonomous drive controller in a conventional vehicle. Since the data to be transmitted in the autonomous driving is not only a short emergency message but also a large amount of image or image data, it is practically impossible to encrypt all related data. Although it may be difficult to encrypt image data in an autonomous drive controller in a vehicle, the offloading technique on a heterogeneous network does not sufficiently provide such a cryptographic service.

Offloading technology typically performs a service program on another device with much better performance, and only receives the result [52]. Computing offloading may be implemented using an RSU or a base station [53-55]. The application of data offloading technology is as follows. Among the plurality of vehicles in the security cluster, there is a leader vehicle that communicates with the BS (Base Station)/RSU (Road Side Unit). The leader vehicle can perform the data encryption operation using an offloading technique. When data encryption is required, the associated data can be transmitted to the base station or the RSU. MEC (Mobile Edge Computing) platform installed in the base station or RSU can provide data encryption services.

*C.* Geo-Fence for Safe Zone

While performing communications within a secure cluster, any vehicle may have a significantly reduced secrecy capacity. What should I do? The geofence can perform the function of notifying the vehicle leaving the security cluster [56]. Generally, a geofence is a system that announces a departure from a region of interest. Depending on the functionality of the geofence, the host vehicle may perform additional countermeasures corresponding to the departure of the secure cluster. A method may be sought to find an alternative to a vehicle that has been displaced, or to increase secrecy capacity by recalling a vehicle that has left the vehicle. Since it is difficult to implement a geofence on a host vehicle, the geofence may be implemented in a nearby RSU / BS using offloading techniques.

In addition, techniques to apply beam-forming [57,58], vary a power [59-61], appropriately select relays [62-64], and generate of artificial noise or jamming [65-66] should be used to enhance physical layer security. On the other hand, physical layer security can be applied as a primary security method in vehicle communication. A supplemental security mechanism for the physical layer security is the conventional application security used in hybrid vehicle communications. The opposite may also prove fruitful as vehicle communication security has not yet become standardized, as well as the fact that this technology uses a physical layer security as a supplement to traditional application security. In depth research in this area will become of increasing focus in the future since using physical layer security for vehicle communication is a very attractive option, and how to define secrecy capacity in the physical layer security of vehicle communication is a very critical issue. Throughout this paper, we have defined the overall parameters of physical layer security related to vehicle communication and propose vehicle communication accordingly. This paper offers a brief glimpse into an ocean of untapped research domains that will most certainly continue to attract interest in the industry.

Also security was called a code. The code may be primarily programming code, but we understand that as a legal term security means to be achieved within a legal framework. The law for autonomous driving will soon be implemented, and our research believes that it will bring great benefits in achieving security in autonomous driving. We hope that this study will help us be able to run freely and safely in the coming future era.

Our research is just the beginning of physical layer security in autonomous driving. In particular, the main research has been done on physical layer security using WAVE. However, it is known that physical layer security research using LTE is also active. In LTE-based vehicle communication, there is a need for further research on physical layer security. In the next sixth generation of telecommunications, the results of this research on physical layer security need to be established as standards.

VII. SIMULATION RESULTS

The intersection simulation attempted to perform communication between the host and target vehicle but also examined the change in the channel capacity as a function of speed. In the simulation environment, the host and the target vehicle trajectories were manually input into various cases with the center coordinates of the intersection as the origin to examine the change in the channel capacity when each vehicle moves at a random speed. In this case, the vehicle speed is designed to conform to the speed limit of 35 Km/h based on the urban environment. For a simple simulation, we applied the change in SNR value to the distance based on the Lehigh model. We simulated six situations using MATLAB as illustrated in Table 1.

Table 1. Cases of communication at intersections.

| Case | Intersection Diagram | Channel Capacity in 2-D | Channel Capacity in 3-D |
|---|---|---|---|
| Case 1 | 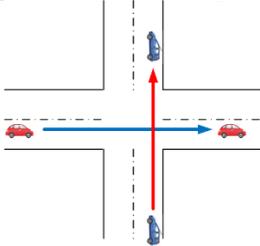 | 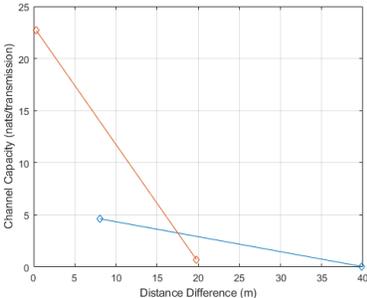 | 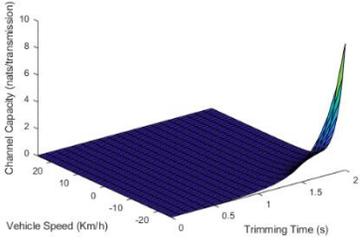 |

Case 2

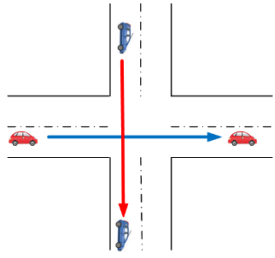 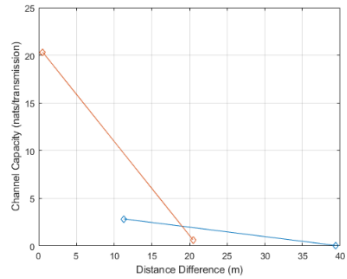 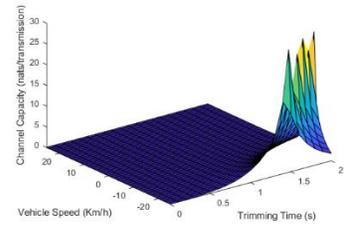

Case 3

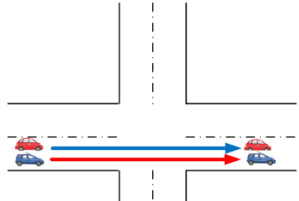 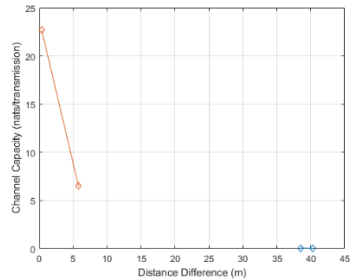 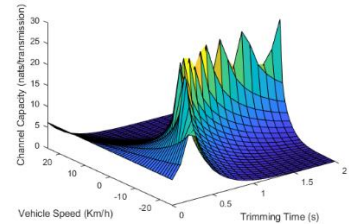

Case 4

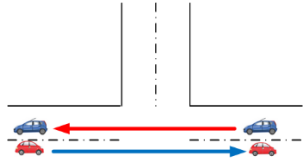 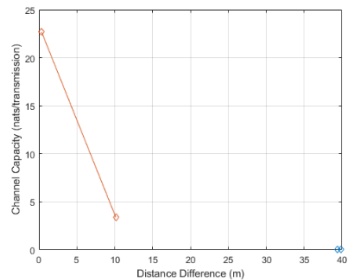 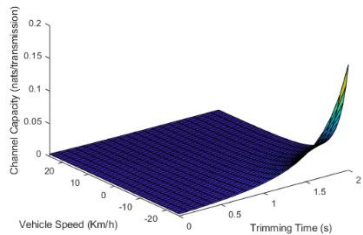

Case 5

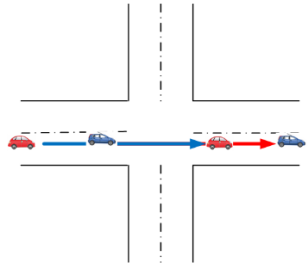 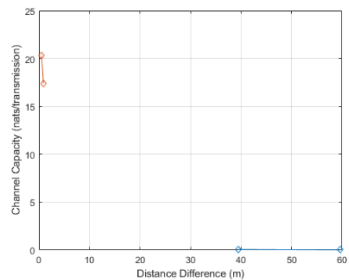 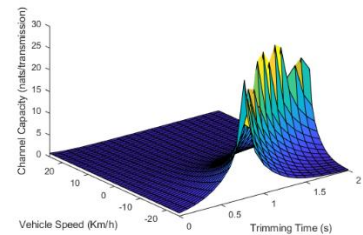

Case 6

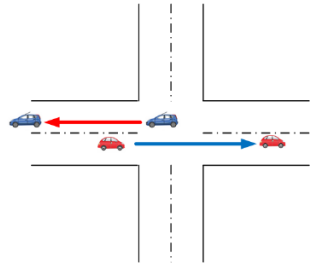 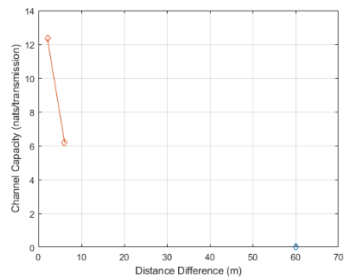 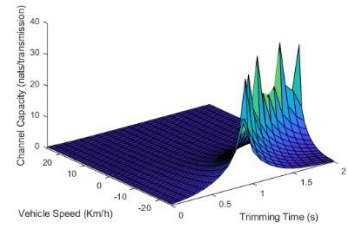

The blue solid line indicates the trajectory of the host vehicle while the red solid line indicates the trajectory of the target vehicle. The trajectory of the host vehicle was set to travel from –60 m to +40 m on the grid. Case 1 indicates communication with a target vehicle moving in a direction orthogonal to the host vehicle. The target vehicle moves from bottom to top. The trajectory of the target vehicle is usually set at –20 m to +20 m. Case 2 is the opposite of Case 1. Case 3 creates an intersection but the target vehicle moves in the same direction as the host vehicle. Case 4 is the opposite direction only of case 3, i.e., the target vehicle moves opposite to the host vehicle. Case 5 is where the host vehicle follows the target vehicle. Case 6 is where the host vehicle follows the target vehicle moving in the opposite direction. We were able to confirm the relationship between the channel capacity and difference in the distance between the host and target vehicle. Larger differences in the distance resulted in a smaller channel capacity of the corresponding vehicle. When the difference in the distance was equal to or larger than the predetermined value, the channel capacity was nearly zero. As illustrated in Table 1, a vehicle's decreasing speed can increase the channel capacity. Summarizing the simulation results, there was a general reduction in the channel capacity as a function of speed. If the distance

Pages **21** / 24

between vehicles was more than a certain value, the channel capacity was nearly zero. Furthermore, if the vehicle speed decreased in the traveling direction, the channel capacity may, to a certain extent, increase. We confirmed that vehicle speed was also a variable parameter to control the secrecy capacity.

In the highway environment, simulation implements movement between moving source vehicle nodes and other vehicle nodes. The simulation conditions are as follows: the road is six lanes, the width of each road is 10m, the simulation distance is 2500 m, the number of source vehicles is 2, the range of the OBU (On-Board Unit) is 2500 m and speed of the vehicle is between 0 and 120 Km /h. And the number of nodes is 25. The total simulation time is 100 seconds. Speed of the vehicle node is arbitrarily changed so that the cluster is formed according to the distance between the source vehicle and the vehicle node. Find the node with the shortest distance from the source vehicle and set the target node as the vehicle node. Since speed of the vehicle node is arbitrary, the clusters have changed from time to time every time the simulation is performed. The following is an exemplary illustration of vehicle node speeds. To verify that the change in secrecy capacity and source vehicle speed is related, we attempted a simulation that forcedly changes speed of the source vehicle. To simplify the related simulation, increasing speed of the source vehicle has been substituted to advance the position of the source vehicle by 5m in the direction of travel, and reducing speed of the source vehicle has reduced the position of the source vehicle by 5m.

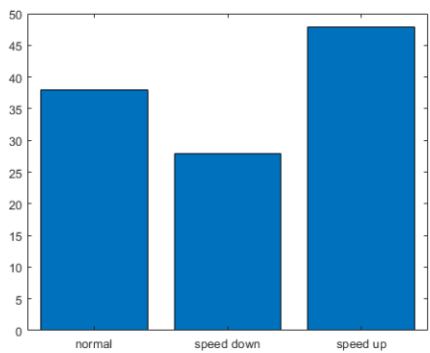

Figure 26. Simulation Result of Distance Relation

Referring to Fig. 26, it can be seen that increasing speed of the source vehicle causes the distance between the source vehicle and the target vehicle to be relatively far apart and the distance between the source vehicle and the target vehicle to be relatively close when speed of the source vehicle is reduced. Where the target vehicle is the vehicle corresponding to the closest distance between the source vehicle and the vehicle nodes. The closer the distance between the source vehicle and the target vehicle is, the greater secrecy capacity is. Conversely, as the distance between the source vehicle and the target vehicle increases, secrecy capacity decreases as the distance increases. However, still considering Doppler effect, increasing speed of the vehicle is expected to increase secrecy capacity. It may be intuitive but somewhat right. This is in contrast to secrecy capacity being inversely proportional to vehicle speed when based on the vehicle speed-related safety distance. Therefore, there is a high possibility that secrecy capacity will be determined in a complex manner in an actual communication environment. This can be the reason why we cannot define the proposed secrecy capacity for vehicle in the vehicle communication using physical layer security.

## VIII. CONCLUSION

In this study, we applied secrecy capacity in vehicle communications. We have found that there are certain limitations in vehicle communications, as compared to applying secrecy capacity in existing wireless communications. These constraints directly or indirectly affect secrecy capacity. For example, speed of a vehicle is closely related to the safety distance in autonomous driving. As the safety distance is closely related to the channel capacity, the result is a close relationship between the vehicle speed and secrecy capacity. Various parameters related to secrecy capacity of vehicle communications were investigated. These parameters made, it possible to confirm that secrecy capacity is controllable. We have also defined vehicular secrecy capacity with only SNR values, and proposed secure V2V communications based on vehicular secrecy capacity. Vehicle communications using secrecy capacity basically adopt the concept of security cluster. The security cluster is defined as vehicle nodes, each having vehicular secrecy capacity greater than a certain value. If you are going to drive a vehicle autonomously in the future, secure communications will be more important than ever, and we are confident that this research will help us achieve secure physical layer security. In the future, actual vehicle communications will be performed from a plurality of transmission antennas and a plurality of reception antennas in the vehicle, and research on this will continue to be conducted. Researches into concrete and diverse techniques for defining security clusters in terms of the outsourcing can be very useful when studies are conducted on, for example, the configuration of security clusters through block chains.

We are confident that our research will be a necessity for implementation of autonomous driving. In addition to the convenience of autonomous driving, the difficulty of security is compelling. In particular, the advent of quantum computers makes it difficult to achieve the security of vehicle communication with the security of existing application layers. At this point, our research will be the first step towards safe vehicle communications from these threats, and we believe that more advanced studies will be made in the future.

## ACKNOWLEDGMENTS

We would like to thank Dr. S.J. Oh for giving us valuable advice early in this research. This work was supported by Samsung Research Funding & Incubation Center for Future Technology under Project Number SRFC-TB1403-51.

## REFERENCES

1. National Highway Traffic Safety Administration (NHTSA), "Federal Motor Vehicle Safety Standards; V2V Communications ," https://www.federalregister.gov/documents/2017/01/12/2016-31059/federal-motor-vehicle-safety-standards-v2v-communications, 2017.
2. L. Sun and Q. Du, "Secure Data Dissemination for Intelligent Transportation Systems," in Secure and Trustworthy Transportation Cyber-Physical Systems, Springer Briefs in Computer Science, pp. 99–140, 2017.
3. F. Saki and S. Sen, "A survey of attacks and detection mechanisms on intelligent transportation systems: VANETs and IoV," Ad Hoc Networks, vol. 61, 2017.
4. F. Camacho, C. Cárdenas and D. Muñoz, "Emerging technologies and research challenges for intelligent transportation systems: 5G, HetNets, and SDN", Int J Interact Des Manu, Springer, 2017.
5. W. Whyte, A. Weimerskirch, V. Kumar and T. Hehn, "A Security Credential Management System for V2V Communications," Vehicular Networking Conference (VNC), 2013 IEEE, 2013.
6. A. Lei, H. Cruickshank, Y. Cao, P. Asuquo, C. P. A. Ogah and Z. Sun, "Blockchain-Based Dynamic Key Management for Heterogeneous Intelligent Transportation Systems," IEEE Internet of Things Journal, vol. 4, no. 6, Dec. 2017.
7. L. Liang, H. Peng, G. Y. Li, and X. S. Shen, "Vehicular Communications: A Physical Layer Perspective," IEEE Transactions on Vehicular Technology, 2017.
8. D. Han, B. Bai and W. Chen, "Secure V2V Communications via Relays: Resource Allocation and Performance Analysis," IEEE Wireless Communications Letters, vol. 6, no. 3, June 2017.
9. M. E. Eltayeb, J. Choi, T. Y. Al-Naffouri and R. W. Heath, Jr., "Enhancing Secrecy with Multi-Antenna Transmission in Millimeter Wave Vehicular Communication Systems," IEEE Transactions on Vehicular Technology, 2017.
10. N.Y. Ahn, D.H. Lee D and S.J. Oh, "Vehicle Communication Using Secrecy Capacity," Advances in Intelligent Systems and Computing, vol 881. Springer, Cham, 2019.
11. F. A. Mullakkal-Babu, M. Wang, B. Arem, R. Happee, "Design and Analysis of Full Range Adaptive Cruise Control with Integrated Collision Avoidance Strategy," 2016 IEEE 19th International Conference on Intelligent Transportation Systems (ITSC), 2016.
12. S. Duan, and J. Zhao, "A Model Based on Hierarchical Safety Distance Aigorithm for ACC Control Mode Switching Strategy," 2017 2nd International Conference on Image, Vision and Computing, 2017.
13. G. Wang, L. Zhao, Y. Hao and J. Zhu, "Design of Active Safety Warning System for Hazardous Chemical Transportation Vehicle," Information Technology and Intelligent Transportation Systems, Springer, pp. 11–21, 2017.
14. D. Zhao, Z. Xia, and Q. Zhang, "Model-free Optimal Control based Intelligent Cruise Control with Hardware-in-the-loop Demonstration," IEEE Computational Intelligence Magazine, May 2017.
15. Jiangfeng Wang, Xuedong Yan, Shuo Nie, and Xiaomeng Li, "Research on Intelligent Vehicle Collision Warning Model Based on Intervehicle Communication," Mathematical Problems in Engineering, Volume 2013, Article ID 208603, 7 pages, 2013.
16. M. Bloch, J. Barros, M. R. D. Rodrigues, and S. W. McLaughlin, "Wireless Information-Theoretic Security," IEEE Transactions on Information Theory, vol. 54, no. 6, June 2008.
17. N.Y. Ahn and D.H. Lee, "Vehicle Secrecy Parameters for V2V Communications" [Online First], IntechOpen, DOI: 10.5772/intechopen.89176. Available from: https://www.intechopen.com/online-first/vehicle-secrecy-parameters-for-v2v-communications, 2019.
18. Y. Zou, J. Zhu, X. Wang, and V. C. M. Leung, "Improving Physical-Layer Security in Wireless Communications Using Diversity Techniques," IEEE Network, January/February 2015.
19. X. Chen, D. W. Kwan Ng, W. H. Gerstacker, and H. Chen, "A Survey on Multiple-Antenna Techniques for Physical Layer Security," IEEE Communications Surveys & Tutorials, vol. 19, no. 2, second quarter 2017.
20. Y. Wu1, W. Liu, S. Wang, W. Guo and X. Chu, "Network Coding in Device-to-device (D2D) Communications Underlaying Cellular Networks," IEEE ICC 2015–Wireless Communication Symposium, 2015.
21. F. Jameel, Faisal, M. A. A. Haider, and A. A. Butt, "Performance Analysis of VANETs under Rayleigh, Rician, Nakagami and Weibull Fading," 2017 International Conference on Communication, Computing and Digital Systems (C-CODE), 2017.
22. H. Inaltekin, M. Chiang, H. V. Poor, and S. B. Wicker, "On Unbounded Path-Loss Models: Effects of Singularity on Wireless Network Performance," IEEE Journal on Selected Areas in Communications, vol. 27, no. 7, Sep. 2009.
23. B.M. ElHalawany, A.A.A. El-Banna, and K. Wu, "Physical-Layer Security and Privacy for Vehicle-to-Everything," IEEE Communications Magazine, Vol. 57 , Issue: 10 , October 2019.
24. A. Zhang and X. Lin, "Security-Aware and Privacy-Preserving D2D Communications in 5G," IEEE Network, July/Aug 2017.
25. A. Zhang, L. Wang, X. Ye and L. Zhou, "Secure content delivery over device-to-device communications underlaying cellular networks," Wireless Communications and Mobile Computing, vol. 27, June 2016.
26. S. Iwata, T. Ohtsuki, and P. T. Kam, "A Lower Bound on Secrecy Capacity for MIMO Wiretap Channel Aided by a Cooperative Jammer with Channel Estimation Error," IEEE Access, vol. 5, 2017.
27. Y. Liang, H. V. Poor, and S. Shamai, "Secure Communication Over Fading Channels," IEEE Transactions on Information Theory, vol. 54, no. 6, June 2008.
28. H. Vincent Poora, and Rafael F. Schaeferb, "Wireless physical layer security," PNAS, vol. 114, no. 1, pp. 19–26, Jan. 2017.
29. M. Ghogho and A. Swami, "Physical-Layer Secrecy Of MIMO Communications In The Presence Of A Poisson Random Field Of Eavesdroppers," Communications Workshops (ICC), 2011 IEEE International Conference on, 2011.
30. G. Chen, J. P. Coon, and M. D. Renzo, "Secrecy Outage Analysis for Downlink Transmissions in the Presence of Randomly Located Eavesdroppers," IEEE Transactions on Information Forensics and Security, vol. 12, no. 5, May 2017.
31. J. Tang, G. Chen, and J. P. Coon, "The Meta Distribution of secrecy Rate in the Presence of Randomly Located Eavesdroppers," IEEE Wireless Communications Letters, vol. PP, no. 99, 2018.

32. X. Ge, H. Cheng, G. Mao, Y. Yang, and S. Tu, "Vehicular Communications for 5G Cooperative Small-Cell Networks," IEEE Transactions on Vehicular Technology, vol. 65, no. 10, Oct. 2016.
33. G. Shiqi, X. ChengWen, F. ZeSong and K. JingMing, "Cooperative Beamforming Design for Physical-Layer Security of Multi-Hop MIMO Communications" Science China Information Sciences, 2015.
34. D. Xu, P. Ren, Q. Du and L. Sun, "Hybrid secure beamforming and vehicle selection using hierarchical agglomerative clustering for C-RAN-based vehicle-to-infrastructure communications in vehicular cyber-physical systems," International Journal of Distributed Sensor Networks, 2016.
35. T. Amin, "Performance Analysis of Secondary Users in Heterogeneous Cognitive Radio Network," Georgia Southern University Digital Commons@Georgia Southern, 2016.
36. C. Chen H. Xiang T. Qiu C. Wang Y. Zhou V. Chang "A rear-end collision prediction scheme based on deep learning in the Internet of Vehicles" J. Parallel Distrib. Comput. vol. 117 pp. 192-204 2018.
37. Z. Lu, W. Liu, Q. Wang, G. Qu and Z. Liu, "A Privacy-Preserving Trust Model Based on Blockchain for VANETs," IEEE Access, vol. 6, pp. 45655-45664, 2018.
38. Awais Hassan, M., Habiba, U., Ghani, U., & Shoaib, M. (2019). A secure message-passing framework for inter-vehicular communication using blockchain. International Journal of Distributed Sensor Networks. https://doi.org/10.1177/1550147719829677
39. Y. Dai, D. Xu, S. Maharjan, Z. Chen, Q. He and Y. Zhang, "Blockchain and Deep Reinforcement Learning Empowered Intelligent 5G Beyond," IEEE Network, vol. 33, no. 3, pp. 10-17, May/June 2019.
40. Y. Yang, L. Chou, C. Tseng, F. Tseng and C. Liu, "Blockchain-Based Traffic Event Validation and Trust Verification for VANETs," IEEE Access, vol. 7, pp. 30868-30877, 2019.
41. J. A. Leon Calvo and R. Mathar, "Secure Blockchain-Based Communication Scheme for Connected Vehicles," 2018 European Conference on Networks and Communications (EuCNC), Ljubljana, Slovenia, 2018, pp. 347-351.
42. Yang, Liuqing et al., "Vehicle-to-vehicle communication based on a peer-to-peer network with graph theory and consensus algorithm," IET Intelligent Transport Systems(2019), 13 (2):280
43. J.T. Oh, K.Y. Kim, and J.Y. Park, "Method for selecting consensus node using nonce and method and apparatus for generating blockchain using the same," US Patent App. 16/389,620, 2019.
44. A. Orsdemir, H. O. Altun, G. Sharma, and M. F. Bocko, "On the Security and Robustness of Encryption via Compressed Sensing," Military Communications Conference (MILCOM), San Diego, CA, USA, Nov. 17-19, 2008, pp. 1–7.
45. V. Athira, S. N. George, and P. P. Deepthi, "A Novel Encryption Method Based on Compressive Sensing," 2013 International Mutli-Conference on Automation, Computing, Communication, Control and Compressed Sensing (iMac4s), 2013.
46. R. Dautov and G. R. Tsouri, "Establishing Secure Measurement Matrix For Compressed Sensing Using Wireless Physical Layer Security," 2013 International Conference on Computing, Networking and Communications, Communications and Information Security Symposium, 2013.
47. J. E. Barceló-Lladó, A. Morell, and G. Seco-Granados, "Amplify-and-Forward Compressed Sensing as a Physical-Layer Secrecy Solution in Wireless Sensor Networks," IEEE Transactions on Information Forensics and Security, vol. 9, no. 5, May 2014.
48. C. H. Lin, S. H. Tsai and Y. P. Lin, "Secure MIMO Transmission via Compressive Sensing," IEEE ICC 2015 - Communication and Information Systems Security Symposium, 2015.
49. J. Choi, "Secure Transmissions via Compressive Sensing in Multicarrier Systems," IEEE Signal Processing Letters, vol. 23, no. 10, Oct. 2016.
50. J. Guo, B. S. Y. He, and F. R. Yu, "A Survey on Compressed Sensing in Vehicular Infotainment Systems," IEEE Communications Surveys & Tutorials, vol. 19, no. 4, fourth quarter 2017.
51. W. Cho, and N. Y. Yu, "Secure Communications With Asymptotically Gaussian Compressed Encryption," IEEE Signal Processing Letters, vol. 25, no. 1, Jan. 2018.
52. Y. Wu et al., "Secrecy-Driven Resource Management for Vehicular Computation Offloading Networks," IEEE Network, vol. 32, no. 3, pp. 84-91, May/June 2018.
53. D. Xu et al., "A Survey of Opportunistic Offloading," IEEE Communications Surveys & Tutorials, vol. 20, no. 3, pp. 2198-2236, third quarter 2018.
54. S. Raza S. Wang M. Ahmed M. R. Anwar "A Survey on Vehicular Edge Computing: Architecture Applications," Technical Issues and Future Directions vol. 2019, 2019.
55. S. S. Shah, M. Ali, A. W. Malik, M. A. Khan and S. D. Ravana, "vFog: A Vehicle-Assisted Computing Framework for Delay-Sensitive Applications in Smart Cities," IEEE Access, vol. 7, pp. 34900-34909, 2019.
56. Yannick Oskar Scherr, Bruno Albert Neumann Saavedra, Mike Hewitt and Dirk Christian Mattfeld, "Service network design with mixed autonomous fleets," Transportation Research Part E: Logistics and Transportation Review, ISSN: 1366-5545, Vol: 124, Page: 40-55, 2019.
57. Y. Wu, A. Khisti, C. Xiao, G. Caire, K. Wong and X. Gao, "Guest Editorial Physical Layer Security for 5G Wireless Networks, Part I," IEEE Journal on Selected Areas in Communications, vol. 36, no. 4, pp. 675-678, April 2018.
58. F. Zhu, F. Gao, H. Lin, S. Jin, J. Zhao and G. Qian, "Robust Beamforming for Physical Layer Security in BDMA Massive MIMO," IEEE Journal on Selected Areas in Communications, vol. 36, no. 4, pp. 775-787, April 2018.
59. R.H. Khokhar, T. Zia, K.Z. Ghafoor, J. Lloret, M. Shiraz, "Realistic and efficient radio propagation model for V2X communications," KSII Transactions on Internet and Information Systems 7 (8), 1933-1954. 2013.
60. T. Liu, S. Han, W. Meng, C. Li and M. Peng, "Dynamic power allocation scheme with clustering based on physical layer security," IET Communications, Volume 12, Issue 20, p. 2546-25551, December 2018.
61. G. Chen, Y. Zhan, Y. Chen, L. Xiao, Y. Wang and N. An, "Reinforcement Learning Based Power Control for In-Body Sensors in WBANs Against Jamming," in IEEE Access, vol. 6, pp. 37403-37412, 2018.
62. K.Z. Ghafoor, K. Abu Bakar, J. Lloret, R.H. Khokhar, K.C. Lee, "Intelligent beaconless geographical forwarding for urban vehicular environments," Wireless networks 19 (3), 345-362, 2013.
63. Y. Huo, Y. Tian, L. Ma, X. Cheng and T. Jing, "Jamming Strategies for Physical Layer Security," IEEE Wireless Communications, vol. 25, no. 1, pp. 148-153, February 2018.
64. L. Hu et al., "Cooperative Jamming for Physical Layer Security Enhancement in Internet of Things," IEEE Internet of Things Journal, vol. 5, no. 1, pp. 219-228, Feb. 2018.
65. Y. Zhou et al., "Improving Physical Layer Security via a UAV Friendly Jammer for Unknown Eavesdropper Location," IEEE Transactions on Vehicular Technology, vol. 67, no. 11, pp. 11280-11284, Nov. 2018.
66. R. Hussain, F. Hussain, and S. Zeadally , "Integration of VANET and 5G Security: A review of design and implementation issues," Citation DataFuture Generation Computer Systems, ISSN: 0167-739X, Vol: 101, Page: 843-864, 2019.